\documentclass[journal,12pt,onecolumn]{IEEEtran}
\usepackage{amsmath,amssymb, dsfont,amssymb}
\usepackage{epsfig,latexsym,amssymb,amsmath,graphics}
\usepackage{graphicx,subfigure}
\usepackage[linesnumbered,ruled]{algorithm2e}
\usepackage{cite,color}
\usepackage{url}

\newtheorem{prop}{Proposition}

\newtheorem{cor}{Corollary}

\newtheorem{lm}{Lemma}

\newtheorem{thm}{Theorem}

\newcommand{\be}{\begin{eqnarray}}
\newcommand{\ee}{\end{eqnarray}}
\newcommand{\benn}{\begin{eqnarray*}}
\newcommand{\eenn}{\end{eqnarray*}}
\def\IR{\rm I \kern-0.20em R}

\newcommand{\bthm}{\begin{thm}}
\newcommand{\ethm}{\end{thm}}

\newcommand{\bcor}{\begin{cor}}
\newcommand{\ecor}{\end{cor}}
\newcommand{\bprop}{\begin{prop}}
\newcommand{\eprop}{\end{prop}}
\newcommand{\blm}{\begin{lm}}
\newcommand{\elm}{\end{lm}}
\newcommand{\beq}{\begin{equation}}
\newcommand{\eeq}{\end{equation}}
\newcommand{\ber}{\begin{eqnarray}}
\newcommand{\eer}{\end{eqnarray}}

\newcommand{\bproof}{\begin{proof}}
\newcommand{\eproof}{\end{proof}}

\newcommand{\bit}{\begin{itemize}}
\newcommand{\eit}{\end{itemize}}
\newcommand{\ben}{\begin{enumerate}}
\newcommand{\een}{\end{enumerate}}
\newcommand{\bdesc}{\begin{description}}
\newcommand{\edesc}{\end{description}}
\newcommand{\beqarrn}{\begin{eqnarray*}}
\newcommand{\eeqarrn}{\end{eqnarray*}}
\newcommand{\bproofof}{\begin{proofof}}
\newcommand{\eproofof}{\end{proofof}}
\newenvironment{rem}{\begin{trivlist}\item[]{\bf
Remark:}\hspace{4mm}}{\end{trivlist}}
\newcommand{\brem}{\begin{rem}}
\newcommand{\erem}{\end{rem}}
\newenvironment{rems}{\begin{trivlist}\item[]{\bf
Remarks}\begin{itemize}}{\end{itemize}\end{trivlist}}
\newcommand{\brems}{\begin{rems}}
\newcommand{\erems}{\end{rems}}
\newtheorem{fact}{Fact}
\newcommand{\bfact}{\begin{fact}}
\newcommand{\efact}{\end{fact}}
\newtheorem{examp}{Example}
\newcommand{\bexamp}{\begin{examp}\rm}
\newcommand{\eexamp}{\end{examp}}
\newtheorem{defn}{Definition}
\newcommand{\bdefn}{\begin{defn}\rm}
\newcommand{\edefn}{\end{defn}}

\newtheorem{alg}{Algorithm}
\newcommand{\balg}{\begin{alg}}
\newcommand{\ealg}{\end{alg}}

\newtheorem{prob}{Problem}
\newcommand{\bprob}{\begin{prob}}
\newcommand{\eprob}{\end{prob}}

\newcommand{\bvtm}{\begin{verbatim}}
\newcommand{\bfig}{\begin{figure}}
\newcommand{\efig}{\end{figure}}
\newcommand{\bcen}{\begin{center}}
\newcommand{\ecen}{\end{center}}

\long\def\comment#1{}

\def \n2{{N_0 \over 2}}

\def \h5{\hspace{0.5in}}

\newcommand{\dff}{\stackrel{\triangle}{=}}

\renewcommand{\baselinestretch}{1}

\def\IR{\mathbb R}

\renewcommand{\baselinestretch}{1.6}

\newtheorem{theorem}{Theorem}
\newtheorem{lemma}{Lemma}

\newtheorem{coro}{Corollary}
\newtheorem{proposition}{{Proposition}}

\title{Characterization on Practical Photon Counting Receiver in Optical Scattering Communication}

\author{Difan Zou, Chen Gong, Kun Wang and Zhengyuan Xu
\thanks{This work was supported by National Key Basic Research Program of China (Grant No. 2013CB329201), Key Program of National Natural Science Foundation of China (Grant No. 61631018), Key Research Program of Frontier Sciences of CAS (Grant No. QYZDY-SSW-JSC003), Key Project in Science and Technology of Guangdong Province (Grant No. 2014B010119001), Shenzhen Peacock Plan (No. 1108170036003286), and the Fundamental Research Funds for the Central Universities.}
\thanks{The authors are with Key Laboratory of Wireless-Optical Communications, Chinese Academy of Sciences, School of Information Science and Technology, University of Science and Technology of China, Hefei, Anhui 230027, China. Z. Xu is also with Shenzhen Graduate School, Tsinghua University, Shenzhen, China. Email: \{knowzou,wangkun1\}@mail.ustc.edu.cn,\{cgong821,xuzy\}@ustc.edu.cn.}}

\begin{document}

\maketitle{}

\begin{abstract}
We characterize the practical photon-counting receiver in optical scattering communication with finite sampling rate and electrical noise. In the receiver side, the detected signal can be characterized as a series of pulses generated by photon-multiplier (PMT) detector and held by the pulse-holding circuits, which are then sampled by the analog-to-digit convertor (ADC) with finite sampling rate and counted by a rising-edge pulse detector. However, the finite small pulse width incurs the dead time effect that may lead to sub-Poisson distribution on the recorded pulses. We analyze first-order and second-order moments on the number of recorded pulses with finite sampling rate at the receiver side under two cases where the sampling period is shorter than or equal to the pulse width as well as longer than the pulse width. Moreover, we adopt the maximum likelihood (ML) detection. In order to simplify the analysis, we adopt binomial distribution approximation on the number of recorded pulses in each slot. A tractable holding time and decision threshold selection rule is provided aiming to maximize the minimal Kullback-Leibler (KL) distance between the two distributions.  The performance of proposed sub-Poisson distribution and the binomial approximation are verified by the experimental results. The equivalent arrival rate and holding time predicted by the of sub-Poisson model and the associated proposed binomial distribution on finite sampling rate and the electrical noise are validated by the simulation results. The proposed the holding time and decision threshold selection rule performs close to the optimal one.
\end{abstract}

\thispagestyle{empty}
\section{Introduction}
On some specific occasions where the conventional RF is prohibited and direct link transmission cannot be guaranteed, the non-line-of-sight (NLOS) optical scattering communication provides an alternative solution to achieve certain information transmission rate~\cite{xu2008ultraviolet}. Optical scattering communication is typically studied in the ultraviolet (UV) spectrum due to a solar blind region (200nm-280nm) where the solar background radiation is negligible \cite{xu2008ultraviolet}. On the UV scattering communication channel characterization, extensive studies on the Monte carlo simulation \cite{ding2009modeling,zhang2012charac,xu2015effects,song2016multi}, theoretical analysis \cite{xiao2011non,gupta2012NLOS,zuo2013closed,sun2016closed} and experimental results \cite{chen2014expe,liao2015uv,raptis2016power}, show that the atmospheric attenuation among scattering channel can be extremely large, especially for long-range transmission. Hence, it is difficult to detect the received signals using conventional continuous waveform receiver, such as photon-diode (PD) and avalanche photon-diode (APD). Instead, a photon-counting receiver is widely deployed.

For photon-counting receiver, the received signals are usually characterized by discrete photoelectrons, whose number in a certain interval satisfies a Poisson distribution. For such a Poisson channel, recent works mainly focus on the channel capacity, such as the continuous Poisson channel capacity \cite{wyner1988capacity,frey1991information} , discrete Poisson channel capacity \cite{lapidoth2009capacity,cao2014capacity1,cao2014capacity2}, wiretap Poisson channel capacity \cite{Laourine2012degradedPoisson}, as well as the Poisson interference channel capacity \cite{lai2015capacity}. Besides, the system characterization and optimization, as well as the signal processing \cite{el2012binary,ScatteringSIMO,Elshimy2015spatial,lai2015optimal,gong2015non,song2016multi,ardakani2017relay,ardakani2017performance} have also been extensively studied from the receiver side.

Most information theory and signal processing works assume perfect photon-counting receiver, which is difficult to realize. A practical photon-counting receiver typically consists of a photon-multiplier (PMT) and the subsequent processing blocks~\cite{becker2005advanced}. A solution is that PMT detects the arriving photons and generates a series of pulses, which are detected by the pulse-holding circuit to generate a series of square pulses with certain width, and further detected by the rising-edge detector to count the pulse number. However, the square pulses generated by pulse-holding circuits typically have finite small width that incurs the dead time effect \cite{cherry2012physics}, where a photon arriving during the pulse duration of the previous photon cannot be detected due to the merge of two pulses. Based on such effect, the photon counts may not satisfy a Poisson distribution. The dead time effect and the model of sub-Poisson distribution for the photon-counting processing have been investigated in \cite{Omote1990deadtime,daniel2000mean}, whose variance is lower than its mean. The photon-counting system with dead time effect has been investigated in optical communication for channel characterizations \cite{Drost2015deadtime,sarbazi2015detection}, and experimental implementation~\cite{chitnis2014spad,shentu2013217}. However, these works assume infinite sampling rate, and identical shape for all pulses, which cannot be realized. The finite sampling rate in ADC and shot noise of PMTs incur further loss of photon counting rate, where the distribution of detected pulse numbers needs to be characterized.

In our work, we model the architecture of a practical photon-counting receiver using a PMT, a pulse-holding circuit, an finite sampling rate ADC and a rising-edge detector. We first assume no electrical noise, and analyze the final distribution under two cases: the sampling period shorter than or equal to the pulse width, and the sampling period longer than the pulse width. Based on the first-order and second-order moments estimation, we derived that for practical photon arrival rate, small pulse width and finite sampling rate, the counting process can still be characterized by the sub-Poisson model for both cases. Then we consider the receiver with shot noise assuming no thermal noise. The results on the first-order and second-order moments estimation show that the sub-Poisson model can still well characterize the practical system, where the corresponding parameters can be formulated as functions with respect to the sampling period, pulse width, and shot noise. When the thermal noise is added, based on the first-order and second-order moments of detected pulses we adopt a binomial approximation on the detected signals. Finally, we consider the on-off keying (OOK) modulation and maximum likelihood (ML) detection. To optimize the holding time and decision threshold for the rising-edge detector that minimizes the error probability, we propose a selection rule on the holding time and detection threshold, which aims to maximize the minimal KL distance of two approximated binomial distributions.  Experimental and numerical results validate the effectiveness of sub-Poisson model and evaluate the performance of the proposed holding time and decision threshold selection rule.

The remainder of this paper is organized as follows. In Section II, we propose the model of PMT-based practical photon-counting receiver and present the existing results on the sub-Poisson distribution for finite-sampling rate.  In Section III, we assume finite sampling rate with no electrical noise, and obtain first-order and second-order moments of detected pulses with finite sampling rate. In Sections IV, first-order and second-order moments on the number of detected pulses are addressed under both the shot and thermal noise. Optimizing the circuit holding time and the detection threshold, as well as the associated maximum likelihood signal detection are proposed in Section V. Experimental and numerical results are shown to verify the effectiveness of the sub-Poisson characterization of practical photon-counting receiver, as well as evaluate the performance of the proposed holding time and decision threshold selection rule in Section VI. Finally, we conclude this paper in Section VII.

\section{Practical System Model for discrete Photon-counting}
\subsection{Practical Photon Counting Process}
Consider a practical photon-counting receiver for optical wireless communication, which contains a PMT detector, a pulse-holding circuit, an ADC, and a rising-edge pulse detector. The practical photon-counting receiver architecture is shown in  Figure~\ref{diagram_processing}.

The entire PMT detector architecture comprises of two parts: an photoelectric converter, and a post-amplifier. Upon one photon is received, the PMT detector generates a short continuous pulse; the pulse-holding circuit detects each short pulse and then outputs a square pulse with certain width. The output signal of pulse-holding circuit is sampled by the ADC and then quantized according to a certain threshold. We adopt a rising-edge pulse detector, where one pulse is recorded upon detecting a rising edge from zero to one.
\begin{figure}[htb]
\centering
\includegraphics[width = 1\columnwidth]{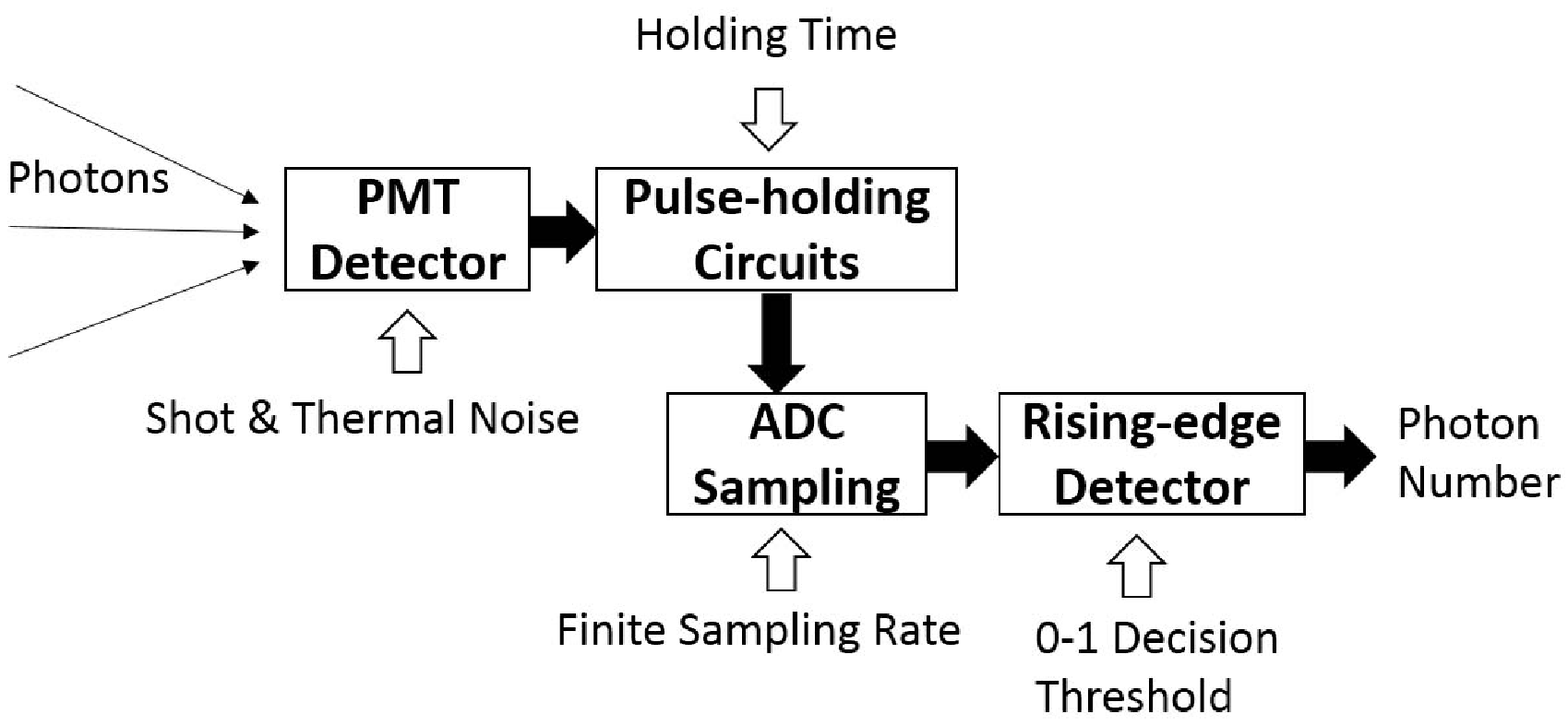}
\caption{The architecture diagram of photon-counting detector.}
\label{diagram_processing}
\end{figure}

\subsection{Signal Model for PMT Detector}
For optical wireless scattering communication, due to the large channel attenuation, the detected optical signal can be characterized as discrete photoelectrons in a symbol duration of length $T_s$. The number of detected photoelectrons, denoted as $N$, satisfies a Poisson distribution. For OOK modulation, let $\lambda_0$ denote the mean number of detected photoelectrons for symbol zero, which is that for the background radiation. Let $\lambda_1=\lambda_s+\lambda_0$ denote the mean number of detected photoelectrons for OOK symbol one, which is the summation of the signal component $\lambda_s$ and background radiation component $\lambda_0$.

We characterize the continuous pulses generated by each detected photoelectron. Let $f(t-t_p)$ denote the square pulse generated by one detected photoelectron by PMT detector and pulse-holding circuit, where $t_p$ denotes the photon arrival time, given by
\be
f(t-t_p)=Ag(t-t_P)+v(t),
\ee
where $A$ denotes the random Gaussian amplitude with mean one due to the shot noise, and $v(t)$ denotes additive Gaussian white thermal noise with mean zero. Note that waveform $g(t)$ depends on the PMT architecture, which is assumed to be known. Let $\sigma^2$ and $\sigma_0^2$ denotes the variances of the $A$ and $v(t)$, respectively, and the thermal noise variance is given by
\be
\sigma_0^2=\frac{2k_eT^0T_s}{R},
\ee
where $k_e$ denotes the Boltzmann constant; $T^0$ denotes the temperature $(K)$; and $R$ denotes the load resistance.

Let $F(t)$ denote the pulse-holding circuits output signal generated by a series of short pulses from PMT, which is sampled by the ADC. Let $F[t_k]$ denote the quantized samples according to the threshold, given as follows,
\be
F[t_k]=\left\{\begin{array}{ll}
0, & F(t_k)<\xi, \\
1, & F(t_k)\ge\xi,
\end{array} \right.
\ee
where $\xi$ denotes the decision threshold. Recall that a photoelectron is recorded upon detecting $0-1$ rising edge. Letting $n[k]$ denote the number of recorded photonelectrons for two samples at $t_k$ and $t_{k+1}$, we have
\be
n[k]=\left\{\begin{array}{ll}
1, & F[t_{k+1}]-F[t_{k}]=1; \\
0, & otherwise.
\end{array} \right.
\ee

\subsection{Distribution of Photon Counting with Dead Time}

Note that the square pulses generated by a practical PMT detector and pulse-holding circuits have certain widths, which enables the pulse detection via finite-rate sampling. However, such pulse width incurs dead time effect that may lead to photon counting loss. When a photon arrives in the dead time duration of the previous photon, the two pulses will merge into one, where only one photoelectron is counted. Such effect is called ``dead time effect", where the duration of photon arrival time leading to the merge of two pulses is denoted as $\tau_0$. In other words, when a photoelectron is detected at the time $t$, a dead time interval from $t$ to $t+\tau_0$ is generated, during which the next arriving photon cannot be recorded.

To simplify the analysis, we normalize the symbol duration interval to $[0,1]$, and the dead time is normalized to $\tau=\tau_0/T_s$. The number of recorded pulses $n$ must be less than the true number of photons $N$. Assuming sufficiently high sampling rate and zero noise variances of the PMT detector, the probability mass function (PMF) of detected photoelectrons number $n$ is given by \cite{Omote1990deadtime}, summarized by the following result.
\begin{proposition}
Given dead time $\tau$ and photon arrival rate $\lambda$, the probability for the number $n$ of detected pulses is given by the following probability function,
\be\label{eq.dis_subpoisson}
\mathbb P(n|\lambda,\tau)=\sum^{M-n}_{m=0}\frac{(-1)^m}{n!m!}\left[\left(1-(n+m-1)\tau\right)\lambda e^{-\lambda\tau}\right]^{n+m},
\ee
where integer $M\dff \lfloor\frac{1}{\tau}\rfloor+1$ defines the maximum number of counted pulses. Moreover,
the mean and variance of $n$ are given as follows,
\be
\mathbb E[n]&=&\lambda e^{-\lambda\tau},\label{eq.true_mean} \\
\mathbb D[n]&=&\mathbb E[n]-\left[1-(1-\tau)^2\right]\mathbb E[n]^2.\label{eq.true_var}
\ee
$\hfill \Box$
\end{proposition}

In general, for the sufficiently short dead time $\tau$, the variance can be approximated by $D[n]\approx\mathbb E[n]-2\tau\mathbb E[n]^2$. According to (\ref{eq.true_mean}) and (\ref{eq.true_var}), the variance of $n$ is lower than the mean, and thus the above distribution shows sub-Poisson characteristics.

\section{The sub-Poisson distribution with finite sampling rate}
Note that Equation (\ref{eq.dis_subpoisson}) provides the distribution of detected photoelectrons in a symbol duration with sufficiently high sampling rate. In this section, we characterize the practical photon-counting receiver and the corresponding sub-Poisson distribution under finite sampling rate. It is interesting to see that different sampling rates may lead to different sub-Poisson distributions on the number of detected pulses. Moreover, we analyze the first-order and second-order moments on the distributions of the detected pulse numbers.

To study the relationship between the sampling rate and the distribution of detected pulse numbers, we first assume no shot noise and no AWGN, which implies identical width and height for the pulses generated by all photoelectrons.  We analyze the distribution of detected pulse numbers the two scenarios, of the sampling periods $T \leq \tau$ and $T > \tau$, and provide its mean and variance.  Assume that $N\dff\frac{1}{T}$ is an integer, such that there are $\frac{1}{T}$ samples in each symbol duration.
\subsection{The Distributions for $T>\tau$}
For sampling interval duration $T>\tau$, one pulse can be detected in interval $[kT,(k+1)T]$ in case of no photon arrival in $[kT-\tau,kT]$ and at least one photon arrives in interval $[(k+1)T-\tau,(k+1)T]$. Thus the probability of one pulse detected in this interval is given by $e^{-\lambda \tau}(1-e^{-\lambda \tau})$. For the total number of photoelectrons detected in one symbol duration, denoted as $n_s$, we have the following results on its first-order and second-order moments.
\begin{theorem}
For the pulse number $n_s$, we have the following results on $n_s$,
\be
\mathbb E[n_s]&=&\frac{e^{-\lambda\tau}\left(1-e^{-\lambda \tau}\right)}{T},  \\
\mathbb E[n_s^2]&=&\mathbb E[n_s]+E[n_s]^2\left[1+2T^2-3T\right].
\ee
\begin{proof}
Please refer to Appendix.A.
\end{proof}
\end{theorem}

Similar to the case of $T\le \tau$, we have the following approximation on the mean and variance of $n_s$, for sufficiently small $\lambda T$ and $\lambda\tau<<1$.
\begin{coro}
We have the following approximation on $\mathbb{E}[n_s]$ and $\mathbb{D}[n_s]$ for sufficiently small $\lambda T$ and $\lambda\tau$
\be
\mathbb E[n_s]&\approx&\frac{\tau\lambda}{T}e^{-\frac{\tau\lambda}{T}\cdot\frac{3T}{2}}; \label{eq.mean_2},\\
\mathbb D[n_s]&=&\mathbb E[n_s^2]-\mathbb E[n_s]^2 \label{eq.var_2}\nonumber \\
&\approx& \mathbb E[n_s]-3T\mathbb E[n_s]^2.
\ee
\begin{proof}
Considering $\tau<T<<\frac{1}{\lambda}$, the mean $\mathbb E[n_s]$ given in Theorem~1 can be approximated by
\be\label{eq.appendA_appro_mean}
\mathbb E[n_s]&=&\frac{e^{-\lambda\tau}(1-e^{-\lambda\tau})}{T}\nonumber \\
&\approx&\frac{1}{T} e^{-\lambda\tau}(\lambda\tau-\frac{\lambda^2\tau^2}{2}) \nonumber \\
&=&\frac{\tau\lambda}{T}e^{-\lambda\tau}(1-\frac{\lambda\tau}{2})\nonumber \\
&\approx&\frac{\tau\lambda}{T}e^{-\frac{\tau\lambda}{T}\cdot\frac{3T}{2}}.
\ee
Then the approximation on $\mathbb D[n_s]$ is given by
\be
\mathbb D[n_s]&=&\mathbb E[n_s^2]-\mathbb E[n_s]^2 \nonumber \\
&=&\mathbb E[n_s]-(3T-2T^2)\mathbb E[n_s]^2 \nonumber \\
&\approx&E[n_s]-3TE[n_s]^2.
\ee
\end{proof}
\end{coro}

Similar to the results (7-8), the distribution on the number of detected pulses under finite sampling rate can also be characterized by the sub-Poisson model, where the equivalent dead time increases from $\tau$ to $\frac{3T}{2}$ and the equivalent photon arrival rate decreases from $\lambda$ to $\frac{\tau\lambda}{T}$.
\subsection{The Distributions for  $T\le\tau$}

Consider one sampling interval $[kT,(k+1)T]$ where the photoelectron detection result $n[k]=1$. Since the event $F[kT]<\xi$ occurs if and only if no photon arrives in $[kT-\tau, kT]$, the probability is given by $\mathbb P\left(F[kT]<\xi\right)=e^{-\lambda\tau}$. The event $F[(k+1)T]\ge\xi$ occurs if and only if there is at least one photon arriving in interval $[kT,(k+1)T]$, where the probability is given by $\mathbb P(n[k]=1)=e^{-\lambda \tau}(1-e^{-\lambda T})$. For the total number of photoelectrons detected in one symbol duration, we have the following results on its first-order and second-order moments.

\begin{theorem}
For the pulse number $n_s$, we have the following results,
\be
\mathbb E[n_s]&=&\frac{e^{-\lambda\tau}\left(1-e^{-\lambda T}\right)}{T},  \\
\mathbb E[n_s^2]&=&\mathbb E[n_s]-E[n_s]^2\big[[(1-(\alpha+1)T)(1-(\alpha+2)T)+2T(1-(\alpha+1)T)\frac{1-e^{-\lambda(T-\sigma)}}{1-e^{-\lambda T}}\big],
\ee
where $\alpha$ is an positive integer given by $\alpha=\left\lfloor\frac{\tau}{T}\right\rfloor$.

\begin{proof}
Please refer to Appendix.B.
\end{proof}
\end{theorem}

Assuming that $\lambda T,\ \lambda\tau<<1$, we have the following approximations on $\mathbb{E}[n_s]$ and $\mathbb{D}[n_s]$.

\begin{coro}
We have the following approximation on $\mathbb{E}[n_s]$ and $\mathbb{D}[n_s]$,
\be
\mathbb E[n_s]&\approx&\lambda e^{-\lambda(\tau+\frac{T}{2})} \label{eq.mean_1},\\
\mathbb D[n_s]&=&\mathbb E[n_s^2]-\mathbb E[n_s]^2 \label{eq.var_1}\nonumber \\
&\approx& \mathbb E[n_s]-2(\tau+\frac{T}{2})\mathbb E[n_s]^2.
\ee

\begin{proof}
Then we consider the case of $T<\tau<<\frac{1}{\lambda}$. According to the Theorem~2, the mean $\mathbb E[n_s]$ can be approximated as follows,
\be\label{eq.appendB_approx_mean}
\frac{e^{-\lambda \tau}(1-e^{-\lambda \tau})}{T}&\approx&\frac{e^{-\lambda \tau}(\lambda T-(\lambda T)^2/2)}{T} \nonumber \\
&=&\lambda e^{-\lambda \tau}(1-\frac{\lambda T}{2}) \nonumber \\
&\approx& \lambda e^{-\lambda (\tau+\frac{T}{2})}.
\ee
The approximation on $\mathbb E\left[\sum_{k\neq l}n[k]n[l]\right]$ is given by
\be
\mathbb E\left[\sum_{k\neq l}n[k]n[l]\right]
&=&\mathbb E[n_s]^2\left[(1-T(2\alpha+3)+(\alpha+2)(\alpha+1)T^2)+2T\frac{1-e^{-\lambda(T-\sigma)}}{1-e^{-\lambda T}}\right] \nonumber \\
&\approx& \mathbb E[n_s]^2\left[(1-T(2\alpha+3))+2T\frac{(\alpha+1)T-\tau}{T}\right] \nonumber \\
&=&\mathbb E[n_s]^2\left[1-2(\tau+\frac{T}{2})\right].
\ee
Then the approximated variance $\mathbb D[n_s]$ is given by
\be\label{eq.appendB_approx_var}
\mathbb D[n_s]&=&\mathbb E[n_s^2]-\mathbb E[n_s]^2 \nonumber \\
&\approx&\mathbb E[n_s]-2(\tau+T/2)\mathbb E[n_s]^2.
\ee
\end{proof}
\end{coro}

Comparing the mean and variance [c.f. (\ref{eq.mean_1}) and (\ref{eq.var_1}), respectively] of $n_s$ with the results given in (7-8), it is seen that for $T<\tau<<\frac{1}{\lambda}$, the finite-rate sampling essentially increases the equivalent dead time from $\tau$ to $\tau+T/2$, while the sub-Poisson distribution can still well describe photon-counting process based on rising edge detection.

\section{The Counting Processing Characterization with Electrical Noises}
Recall that the real PMT receiver may suffer both shot and thermal noise with variances $\sigma^2$ and $\sigma_0^2$, respectively. The performance degradation of such two types of noises needs to be analyzed. Moreover, the optimal decision threshold $\xi^*$ under the two types of noise needs to be determined.

For a practical PMT-based photon-counting receiver, thermal noise is significantly weaker than shot noise and signal power, i.e. $\sigma_0<<\sigma$. In this section, we will investigate the counting performance of the counting receiver first under shot noise first, and then under both shot and thermal noise.
\subsection{The Number of Detected Photoelectrons with Merely Shot Noise}
Assuming no additive thermal noise, we analyze the probability $\mathbb{P}(n_k = 1)$.
Consider the probability that the sample at time $kT$, denoted as $F[kT]$, is lower than the decision threshold $\xi$. Assuming a small $T$ such that the mean number $\lambda T$ of arrival photons in the duration of $T$ is small as well, based on which we have the following analysis on the probability $\mathbb P(F[kT]<\xi)$.

\textbf{Case ${\cal H}_1$:} There is no photon arrival events in time interval $[kT-\tau, kT]$. It is obvious that event $F[kT]<\xi$ must occur, i.e., we have $\mathbb P(F[kT]<\xi|{\cal H}_1)\mathbb P({\cal H}_1)=e^{-\lambda\tau}$. We analyze the following three cases.

\textbf{Case ${\cal H}_2$:} There is one photon arriving in interval $[kT-\tau,kT]$. Based on the Gaussian random characteristics of the amplitude of each pulse, we have that \be
\mathbb P(F[kT]<\xi|{\cal H}_2)\mathbb P({\cal H}_2)=e^{-\lambda\tau}\lambda\tau Q(\frac{1-\xi}{\sigma}),
\ee
where Gaussian tail probability $Q(\cdot)$ is given by
\be
Q(x)=\int_{x}^\infty\frac{1}{\sqrt{2\pi}}e^{-\frac{x^2}{2}}.
\ee

\textbf {Case ${\cal H}_3$:} There are more than one photon arriving in the interval $[kT-\tau, kT]$. The probability is given by
\be
\mathbb P(F[kT]<\xi|{\cal H}_3)\mathbb P({\cal H}_3)&=&\sum_{k=2}^{\infty}\frac{(\lambda \tau)^ke^{-\lambda \tau}}{k!}Q\left(\frac{k-\xi}{\sqrt{k\sigma^2}}\right) \nonumber \\
&<&\lambda^2 \tau^2Q\left(\frac{2-\xi}{\sqrt{2\sigma^2}}\right).
\ee
Note that for small $\lambda T$ and $\sigma$, probability $\mathbb P(F[kT]<\xi|{\cal H}_3)\mathbb P({\cal H}_3)$ is significantly lower than that of other two cases. Thus we assume that $\mathbb P(F[kT]<\xi|{\cal H}_3)\mathbb P({\cal H}_3)$ is negligible and can be omitted in the following analysis. More specifically, the following approximation on $\mathbb P(F[kT]<\xi)$ is given by
\be\label{eq.noisecase1_prob1}
\mathbb P(F[kT]<\xi)
&=&\mathbb P(F[kT]<\xi|{\cal H}_1)\mathbb P({\cal H}_1)+\mathbb P(F[kT]<\xi|{\cal H}_2)\mathbb P({\cal H}_2)+\mathbb P(F[kT]<\xi|{\cal H}_3)\mathbb P({\cal H}_3)\nonumber \\
&=&e^{-\lambda \tau}\left[1+\lambda \tau q\right]+\sum_{k=2}^{\infty}\frac{(\lambda T)^ke^{-\lambda \tau}}{k!}Q\left(\frac{k-\xi}{\sqrt{k\sigma^2}}\right)\nonumber\\
&\approx& e^{-\lambda \tau}\left[1+\lambda \tau q\right],
\ee
where $q\dff Q(\frac{1-\xi}{\sigma})$.

Similarly to the previous Section, we analyze the probability of of detecting a rising edge between two samples at $kT$ and $(k+1)T$, i.e., $F[kT]<\xi$ and $F[(k+1)T]>\xi$. Consider two cases, $T>\tau$ and $T\le\tau$.

\subsubsection{Analysis for case $T>\tau$}
It can be seen that any photon arrival event in $[kT-\tau,kT]$ has no impact on the sample at $(k+1)T$, and thus events $F[kT]<\xi$ and $F[(k+1)>\xi]$ are statistically independent. Note that the probability of the former event has been approximated by Equation (\ref{eq.noisecase1_prob1}), and the probability of the latter one is given by
\be\label{eq.noisecase1_prob2}
\mathbb P(F[(k+1)T]>\xi)
&=&1-\mathbb P(F[(k+1)T]<\xi)\nonumber \\
&=&1-e^{-\lambda \tau}\left[1+\lambda \tau q\right]-\sum_{k=2}^{\infty}\frac{(\lambda T)^ke^{-\lambda \tau}}{k!}Q\left(\frac{k-\xi}{\sqrt{k\sigma^2}}\right)\nonumber\\
&\approx& 1-e^{-\lambda \tau}\left[1+\lambda \tau q\right].
\ee
Similarly, assuming sufficiently $\lambda T$ and $\lambda\tau$, we have the following approximation on the mean and variance of $n_s$.
\begin{coro}
The approximation on $\mathbb{E}[n_s]$ and $\mathbb{D}[n_s]$ are given by
\be
\mathbb E[n_s]&\approx&\frac{(1-q)\lambda\tau}{T}e^{-\frac{(1-q)\tau}{T}\cdot\frac{3T}{2}} \label{eq.mean_3},\\
\mathbb D[n_s]&=&\mathbb E[n_s^2]-\mathbb E[n_s]^2 \label{eq.var_3}\nonumber \\
&\approx& \mathbb E[n_s]-3T\mathbb E[n_s]^2.
\ee
\begin{proof}
Please refer to Appendix.C.
\end{proof}
\end{coro}

From the above results, it is seen that under shot noise and sampling period $T$, the equivalent photon arrival rate is reduced to $\frac{\tau\lambda(1-q)}{T}$, while the equivalent dead time remains the same as that without shot noise.

\subsubsection{Analysis for case $T\le\tau$}
We still calculate the probabilities of $F[kT]<\xi$ and $F[(k+1)T]>\xi$. However, the two cases are not statistically independent. When one photon arrives in interval $[(k+1)T-\tau,kT]$, the samples at $(k+1)T$ may be changed. Assuming small $\lambda T$ and $\sigma$, the probability of $F[(k+1)T]>\xi$ is that of at least one photon arriving. To make the analysis tractable, we adopt the approximation of at most one photon arrival in interval $[kT - \tau, kT]$ based on the assumption of sufficiently small $\tau\lambda$. We analyze the probability of $n[k] = 1$ considering the following three cases:

\textbf {Case} ${\cal E}_1$:  No photon arrives in $[kT-\tau,kT]$. Then the probability of $n[k]=1$ is that of a photon arriving between $kT$ and $(k+1)T$, given by
\be
\mathbb P \left(n[k]=1|{\cal E}_1\right)&=&1-\sum_{k=0}^\infty\frac{\lambda^kT^ke^{\lambda T}}{k!}Q\left(\frac{k-\xi}{\sqrt{k\sigma^2}}\right)\nonumber \\
&\approx&1-e^{-\lambda T}(1+\lambda T q),
\ee
where we denote $Q(-\infty)=1$.

\textbf {Case} ${\cal E}_2$: One photon arrives in $[kT-\tau,(k+1)T-\tau]$, and no photon arrives in $[(k+1)T-\tau,kT]$. Since the photon arrival in $[kT-\tau,(k+1)T-\tau]$ does not affect the sample $F[(k+1)T]$, the probability of $n[k]=1$ is given by
\be
\mathbb P \left(n[k]=1|{\cal E}_2\right)&=&\mathbb P (F[kT]<\xi|{\cal E}_2)\mathbb P(F[(k+1)T]>\xi)\nonumber \\
&\approx& q\left[1-e^{-\lambda T}(1+\lambda T q)\right].
\ee

\textbf {Case} ${\cal E}_3$: No photon arrives in $[kT-\tau,(k+1)T-\tau]$, and one photon arrives in $[(k+1)T-\tau,kT]$. Recalling that the probability $\mathbb P(F[kT]<\xi)$ can be approximated to be zero if more than one photon arrives in $[kT-\tau,kT]$, we have the probability of $n[k]=1$ as follows,
\be
\mathbb P \left(n[k]=1|{\cal E}_3\right)=q\left[1-e^{-\lambda T}\right].
\ee
Thus, we also have the following approximation on probability $\mathbb P[n_s=1]$, given by
\be
\mathbb P[n_s=1]
&=& \mathbb P[n_s=1|{\cal E}_1]\mathbb P[{\cal E}_1]+\mathbb P[n_s=1|{\cal E}_2]\mathbb P[{\cal E}_2]+\mathbb P[n_s=1|{\cal E}_3]\mathbb P[{\cal E}_3] \nonumber \\
&=&e^{-\lambda\tau}(1+\lambda Tq)\left[1-e^{-\lambda T}(1+\lambda T q)\right]
+e^{-\lambda\tau}\lambda(\tau-T)q(1-e^{-\lambda T})\nonumber \\
&=&e^{-\lambda\tau}(1+\lambda \tau q)-e^{-\lambda(\tau+T)}\left[(1+\lambda Tq)^2+\lambda(\tau-T)q\right]\nonumber \\
&\approx& e^{-\lambda\tau}(1+\lambda \tau q)\left[1-e^{-\lambda T}\frac{1+\lambda Tq+\lambda \tau q}{1+\lambda \tau q}\right] \nonumber \\
&\approx& e^{-\lambda\tau}(1+\lambda \tau q)\left[1-e^{-\lambda T}(1+\lambda Tq)\right].
\ee

Moreover, we also have the following approximation on the mean and variance of $n_s$.
\begin{coro}
We have the following approximation on $\mathbb{E}[n_s]$ and $\mathbb{D}[n_s]$,
\be
\mathbb E[n_s]&\approx&\lambda(1-q)e^{\lambda(1-q)(\tau+\frac{\tau}{2})} \label{eq.mean_4},\\
\mathbb D[n_s]&=&\mathbb E[n_s^2]-\mathbb E[n_s]^2 \label{eq.var_4}\nonumber \\
&\approx& \mathbb E[n_s]-2(\tau+\frac{T}{2})\mathbb E[n_s]^2.
\ee
\begin{proof}
Please refer to Appendix.D.
\end{proof}
\end{coro}

Based on the above results on the mean and variance, we have that with electrical noise, the equivalent photon arrival rate is reduced to $(1-q)\lambda$ and the equivalent dead time remains the same as that without shot noise.
\subsection{The Number of Detected Photoelectrons with Both Shot and Thermal Noise}
In the photon-counting system, the electrical thermal noises in the PMT and amplifier are significantly weaker than the shot noise, i.e., $\sigma_0<<\sigma$. Consider one sample at $nT$, when there are $k$ ($k\ge1$) pulses merging at $nT$, the variance of this sample is $k\sigma^2+\sigma_0^2$, where the standard deviation is $\sqrt{k\sigma^2+\sigma_0^2}<\sqrt k\sigma+\frac{\sigma_0^2}{2k\sigma^2}$. Since $\frac{\sigma_0^2}{2k\sigma^2}$ is significantly smaller than $\frac{\sigma_0}{\sigma}$, we assume negligible thermal noise in the sample at $nT$ in case of one pulse arrival event that brings shot noise, and consider the probability that the sample at $nT$ exceeds the decision threshold in case of no photon arrivals in $[nT-\tau, nT]$.

As aforementioned in last subsection, the photon-counting process with shot noise can be approximated as a new photon-counting process with its equivalent photon arrival rate. Comparing with the model without shot noise, it can be found that the shot noise only results in a modified photon arrival rate, i.e., $\lambda'\dff (1-q)\lambda$. Hence the counting process with both shot and thermal noise can be approximated as that without shot noise but under a modified photon arrival rate corresponding to the shot noise. Such approximation performs well according to simulation results. Similar to the former analysis on the moments of detected pulse numbers, we investigate the two cases of $T>\tau$ and $T\le\tau$.

\subsubsection{Analysis for Case $T>\tau$}
We calculate the probability of $F[kT]<\xi$ and $F[(k+1)T]>\xi$. Recall the formula (\ref{eq.noisecase1_prob1}), The probability $\mathbb P \left(F[kT]<\xi\right)$ is given by
\be
\mathbb P \left(F[kT]<\xi\right)&\approx& e^{-\lambda\tau}(1-p)+e^{-\lambda\tau}\lambda\tau q \nonumber \\
&\approx&e^{-\lambda\tau}(1-p)+e^{-\lambda\tau}\lambda\tau q(1-p)\nonumber \\
&\approx&e^{-\lambda'\tau}(1-p),
\ee
where $p\dff Q\left(\frac{\xi}{\sigma_0}\right)$ denotes the probability that the thermal noise signal exceeds the decision threshold $\xi$. Then we have the probability of $n[k]=1$, given by
\be\label{eq.prob_5}
\mathbb P[n[k]=1]&=&\mathbb P[F[kT]<\xi] \mathbb P[F[(k+1)T]>\xi]\nonumber\\
&=&e^{-\lambda'\tau}(1-p)\left[1-e^{-\lambda'\tau}(1-p)\right].
\ee

Moreover, we have the following results on the mean and variance of $n_s$.
\begin{coro}
We have the following approximation on $\mathbb{E}[n_s]$ and $\mathbb{D}[n_s]$,
\be
\mathbb E[n_s]&\cong&\frac{e^{-\lambda'\tau}(1-p)\left[1-e^{-\lambda'\tau}(1-p)\right]}{T} \label{eq.mean_5},\\
\mathbb D[n_s]&=&\mathbb E[n_s^2]-\mathbb E[n_s]^2 \label{eq.var_5}\nonumber \\
&\cong& \mathbb E[n_s]+(2T^2-3T)\mathbb E[n_s]^2.
\ee
\begin{proof}
Since the events $F[kT]<\xi$ and $F[(k+1)T]>\xi$ are statistical independent, the above results can be derived by following the identical procedures as that given in Appendix. A, which is omitted here.
\end{proof}
\end{coro}
\subsubsection{Analysis for Case $T\le\tau$}
Similarly, the counting system with shot noise can be approximated as a new counting system with a modified photon arrival rate. One pulse is recorded in $[kT,(k+1)T]$ if $F[kT]<\xi$ and $F[(k+1)T]>\xi$. The probability of one pulse detected in this interval is given by $e^{-\lambda\tau}(1-p)\left[1-e^{-\lambda T}(1-p)\right]$. Considering the total number of detected pulses, we have the following results on its mean and variance of $n_s$.
\begin{coro}
We have the following results on $\mathbb{E}[n_s]$ and $\mathbb{D}[n_s]$,
\be
\mathbb E[n_s]&\cong&\frac{e^{-\lambda'\tau}(1-p)\left[1-e^{-\lambda'T}(1-p)\right]}{T} \label{eq.mean_6},\\
\mathbb D[n_s]&\cong&\mathbb E[n_s]\big[1+2\sum_{s=2}^\alpha(1-sT)p(1-p)e^{-\lambda'sT}\big]\nonumber \\
&&\ \ \ +\mathbb E[n_s]^2\big[(1-(\alpha+1)T)(1-(\alpha+2)T)-1 \nonumber \\
&&\ \ \ +2T(1-(\alpha+1)T)\frac{1-e^{-\lambda'(T-\delta)}(1-p)}{1-e^{-\lambda'T}(1-p)}\big] \nonumber \\
&\approx&\mathbb E[n_s]\left[1+2(\alpha-1)p\right]+2\mathbb E[n_s]^2\left[-(\tau+T/2)+\frac{p\delta}{\lambda' T+p}\right].\label{eq.var_6}
\ee
\begin{proof}
Please refer to Appendix.E.
\end{proof}
\end{coro}

\section{Signal Detection and System Parameter Optimization}
\subsection{Binomial Approximation on Likelihood Functions}
At the receiver side, let $\mathbb{P}(n|\lambda)$ denote the probability of detecting $n$ pulses under where number of detected photoelectrons satisfies a Poisson distribution with mean $\lambda$.  We adopt the maximum likelihood (ML) detection, given as follows,
\be
Y=\left\{\begin{array}{ll}
1, & \mathbb P(n_s|\lambda_1,\tau)\ge\mathbb P(n_s|\lambda_0,\tau), \\
0, & P(n_s|\lambda_1,\tau)<\mathbb P(n_s|\lambda_0,\tau);
\end{array} \right.
\ee
where $\lambda_0$ and $\lambda_1$ denote the mean number of photons for symbols $0$ and $1$, respectively.

However, the complicated term of $\mathbb P(n|\lambda,\tau)$ may make the analysis on the exact error probability intractable. We resort to the KL distance-based criterion, and adopt binomial approximation with the same mean and variance on the probability $\mathbb P(n_s|\lambda,\tau)$.
\begin{theorem}
Based on the mean and variance of $n_s$ given in Corollary $5$ and Corollary $4$, for the mean number of photoelectrons $\lambda$, the parameters in binomial distribution for the two cases $T>\tau$ and $T\le\tau$ are given as follows.

\begin{itemize}
  \item For $T>\tau$, we have that
    \be\label{eq.binoapprox_para1}
    N&=&\frac{1}{2\tau'},\nonumber \\
    P&=&2\tau'\hat N,
    \ee
  \item For $T\le\tau$, we have that
   \be\label{eq.binoapprox_para2}
    N&=&\frac{1}{2\tau'}\frac{1}{1-\left[\frac{p\delta}{\tau'(\lambda' T+p)}+\frac{(\alpha-1)p}{\hat N\tau'}\right]},\nonumber \\
    P&=&2\tau'\hat N\left\{1-\left[\frac{p\delta}{\tau'(\lambda' T+p)}+\frac{(\alpha-1)p}{\hat N\tau'}\right]\right\},
    \ee
\end{itemize}
where $\hat N\dff \mathbb E[n_s]$ denotes the mean recorded pulse number, $\lambda'=\lambda(1-q)$.  $\tau'=\tau+T/2$ and $\tau'=\frac{3T}{2}$ denotes the equivalent dead time for the cases $T\le\tau$ and $T>\tau$, respectively.

\begin{proof}
For the case of $T>\tau$, we have the mean $\mathbb E[n_s]=\hat N=NP$, and the variance $\mathbb D[n_s]=\hat N-2\tau'\hat N^2=NP(1-P)$. Thus Equation (\ref{eq.binoapprox_para1}) can be directly obtained by solving the above two equations.

For the case of $T\le\tau$, we also write the equations on the mean and variance, given by
\be
NP&=&\hat N, \nonumber \\
NP(1-P)&=& \hat N\left\{1+2(\alpha-1)p + 2\hat N\left[-\tau'+\frac{p\delta}{\lambda' T+p}\right]\right\}.
\ee
Dividing the second equation by the first one, we have the following result on $P$
\be\label{eq.binoappro_P}
P&=&2\tau'\hat N -\frac{2p\delta\hat N}{\lambda'T} -2(\alpha-1)p \nonumber \\
&=&2\tau'\hat N\left\{1-\left[\frac{p\delta}{\tau'(\lambda' T+p)}+\frac{(\alpha-1)p}{\hat N\tau'}\right]\right\}.
\ee
Then we have the following result on $N$
\be\label{eq.binoappro_N}
N=\frac{\hat N}{P}=\frac{1}{2\tau'}\frac{1}{1-\left[\frac{p\delta}{\tau'(\lambda' T+p)}+\frac{(\alpha-1)p}{\hat N\tau'}\right]}.
\ee
\end{proof}
\end{theorem}

Based on Theorem~3, it is seen that for $T > \tau$, the binomial distribution approximations for $\lambda_0$ and $\lambda_1$ have the same length but different probability distributions, i.e.,  $N_1 = N_0$ but $P_1 \neq P_0$; but for $T < \tau$, the binomial distribution approximations for $\lambda_0$ and $\lambda_1$ have the different lengths and different probabilities, i.e.,  $N_1 \neq N_0$ and $P_1 \neq P_0$. For such lengthes and probabilities, the parameters $\lambda'$ and $\tau'$ can be estimated via matching the first and second moments of the distributions.
\subsection{The Decision Threshold and Holding Time Optimization}
The decision threshold $\xi$ and holding time $\tau$ needs to be optimized to improve the pulse-counting performance.

As discussed, we consider the KL distance between two binomial distributions rather than the formulation of total error probability, since the optimization on the exact error probability may be intractable. Two approximated likelihood functions, denoted as $P_1^B\dff \mathbb B(N_1,P_1)$ and $P_0^B\dff \mathbb B(N_0,P_0)$, respectively, have the following KL distances for the case of $T>\tau$,
\be\label{eq.KLD1}
D(P_0^B||P_1^B)&=& N_0\left[P_0\log\frac{P_0}{P_1}+(1-P_0)\log\frac{1-P_0}{1-P_1}\right];\nonumber \\
D(P_1^B||P_0^B)&=&N_0\left[P_1\log\frac{P_1}{P_0}+(1-P_1)\log\frac{1-P_1}{1-P _0}\right].
\ee
and the following KL distances for the case of $T\le\tau$,
\be\label{eq.KLD2}
D(P_0^B||P_1^B)&=&\mathbb E_{n|p_0}\left[\sum_{k=N_1+1}^{N_0} \log \frac{k}{k-n}\right]+N_0(1-P_0)\log\frac{1-P_0}{1-P_1}\nonumber \\
&&\quad+N_0P_0\log\frac{P_0}{P_1}+(N_0-N_1)\log(1-P_1);\nonumber \\
D(P_1^B||P_0^B)&=&\mathbb E_{n|p_1}\left[\sum_{k=N_1+1}^{N_0} \log \frac{k-n}{k}\right]+N_1(1-P_1)\log\frac{1-P_1}{1-P_0}\nonumber \\
&&\quad+N_1P_1\log\frac{P_1}{P_0}+(N_1-N_0)\log(1-P_0).
\ee

According to the Chernoff-Stein Lemma \cite{ElementsInfoTheory}, we pursue the optimal threshold $\xi^*$ that maximizes the minimum of the above two KL distances. More specifically, the optimal threshold $\xi^*$ and holding time $\tau^*$ are defined as
\be\label{eq.opt1}
\{\xi^*,\tau^*\}=\arg\max_{\xi,\tau}\min\left\{D(P_0^B||P_1^B),D(P_1^B||P_0^B)\right\}.
\ee
Note that the exact solution to the above optimization problem is intractable. In the remainder of this subsection, we resort to an approximation but tractable solution, which does not show significant loss in simulations.

Note that the probability $P_1$ is of the same order of $\lambda_1\tau$ when $p$ is sufficiently small, and that $P_0<P_1$ if we try to maintain reliable communication. The expectation of $n$ in (\ref{eq.KLD2}) is much smaller than $N_1$ and $N_0$, and thus each term in the summation is close to each other. Therefore, we have the following approximation on the KL distances in (\ref{eq.KLD2}),
\be
D(P_0^B||P_1^B)&\approx& (N_0-N_1)\log\frac{N_0}{N_0-N_0P_0}+N_0P_0\log\frac{P_0}{P_1} \nonumber \\
&&\quad +N_0(1-P_0)\log\frac{1-P_0}{1-P_1}+(N_0-N_1)\log(1-P_1) \nonumber \\
&=&N_1\log\frac{1-P_0}{1-P_1}+N_0P_0\left[\log\frac{P_0}{P_1}-\log\frac{1-P_0}{1-P_1}\right] ; \nonumber \\
D(P_1^B||P_0^B)&\approx&(N_0-N_1)\log\frac{N_1}{N_1-N_1P_1}+N_1P_1\log\frac{P_1}{P_0}\nonumber \\
&&\quad+N_1(1-P_1)\log\frac{1-P_1}{1-P_0}+(N_1-N_0)\log(1-P_0)\nonumber \\
&=&N_0\log\frac{1-P_1}{1-P_0}+N_1P_1\left[\log\frac{P_1}{P_0}-\log\frac{1-P_1}{1-P_0}\right].
\ee

Note that the optical scattering communication is typically operating in the UV spectrum, where the background radiation is sufficiently small. We can formulate the optimization problem assuming sufficiently small $\lambda_0$. First we have the following lemma.
\begin{lemma}
Let $\tau'_0\dff \frac{3T}{2}$. If $\log \frac{\hat N_1}{\hat N_0}>\frac{1+\frac{\tau'}{\tau_0'}}{1-2\tau'\hat N_1}$, we have approximately $D(P_0^B||P_1^B)<D(P_1^B||P_0^B)$ for both cases of $T>\tau$ and $T\le\tau$.

\begin{proof}
Please refer to Appendix.F.
\end{proof}
\end{lemma}

Note that for sufficiently small $p$ and $\lambda_1\tau'$, the condition $\log \frac{\hat N_1}{\hat N_0}>\frac{1+\frac{\tau'}{\tau_0'}}{1-2\tau'\hat N_1}$ can be easily satisfied in the photon-counting system with sufficiently small background radiation if $\frac{\tau'}{\tau_0'}$ is not large. Based on such assumption, the optimization problem in (\ref{eq.opt1}) can be approximated as follows,
\be\label{eq.opt2}
\{\xi^*,\tau^*\}=\arg\max_{\xi,\tau}D(P_0^B||P_1^B).
\ee

\subsubsection{Optimization on $\tau$}

We first consider the optimization on $\tau$. For the case of $T>\tau$, we have the following result.
\begin{lemma}
Let $\gamma\dff\frac{\hat N_1}{\hat N_0}$. For the case of $T>\tau$, we have the optimal holding time $\tau^*=T$ if $p<\frac{1}{2}-\frac{\log \gamma}{2(\gamma-1)}-\gamma\tau$.

\begin{proof}
We write the derivative of $D(P_0^B||P_1^B)$ with respect to $\tau$ as follows,
\be
\frac{\partial D(P_0^B||P_1^B)}{\partial \tau}=\left[\log\frac{\hat N_0}{\hat N_1}-\log\frac{1-2\tau'\hat N_0}{1-2\tau' \hat N_1}\right]\frac{\partial \hat N_0}{\partial \tau}+\left[-\frac{\hat N_0}{\hat N_1}+\frac{1-2\tau'\hat N_0}{1-2\tau' \hat N_1}\right]\frac{\partial \hat N_1}{\partial \tau}.
\ee

Then we show that the above derivative is larger than zero, which is equivalent to proving the following
\be
\frac{\log\frac{1-2\tau'\hat N_0}{1-2\tau' \hat N_1}-\log\frac{\hat N_0}{\hat N_1}}{\frac{1-2\tau'\hat N_0}{1-2\tau' \hat N_1}-\frac{\hat N_0}{\hat N_1}}\le\frac{\frac{\partial \hat N_1}{\partial \tau}}{\frac{\partial \hat N_0}{\partial \tau}}.
\ee

Considering the function $\frac{\log a- \log b}{a-b}$, which is decreasing with respect to both $a$ and $b$. Then the result can be proved if the following can be proved,
\be
\frac{\log\frac{\hat N_1}{\hat N_0}}{1-\frac{\hat N_0}{\hat N_1}}\le\frac{\frac{\partial \hat N_1}{\partial \tau}}{\frac{\partial \hat N_0}{\partial \tau}}=\frac{\lambda_1'e^{-\lambda_1'\tau}\left[2(1-p)e^{-\lambda_1'\tau}-1\right]}{\lambda_0'e^{-\lambda_0'\tau}\left[2(1-p)e^{-\lambda_0'\tau}-1\right]}.
\ee
Defining $h(x)=\frac{1-(1-p)e^{-\beta x}}{x}$, we have that $\frac{\partial h(x)}{\partial x}=\frac{-1+(1-p)(1+\beta x)e^{-\beta x}}{x^2}\le0$, which implies that
\be\label{eq.opt_tau_ineq1}
\frac{\lambda_1'e^{-\lambda_1'\tau}}{\lambda_0'e^{-\lambda_0'\tau}}>\frac{\left[1-(1-p)e^{-\lambda_1' T}\right]e^{-\lambda_1'\tau}}{\left[1-(1-p)e^{-\lambda_0' T}\right]e^{-\lambda_0'\tau}}=\frac{\hat N_1}{\hat N_0}.
\ee
Moreover, we have $\frac{2(1-p)e^{-\lambda_1'\tau}-1}{2(1-p)e^{-\lambda_0'\tau}-1}>2(1-p)e^{-\lambda_1'\tau-1}>2(1-p)(1-\lambda_1'\tau)-1>1-2p-2\lambda_1'\tau$. Then the inequality (\ref{eq.opt_tau_ineq1}) is satisfied  if the following holds
\be
\log\frac{\hat N_1}{\hat N_0}\le\left(\frac{\hat N_1}{\hat N_0}-1\right)(1-2p-2\lambda_1'\tau),
\ee
which is equivalent to the condition given in this lemma.
\end{proof}
\end{lemma}

Note that for sufficiently small background radiation $\lambda_0$, $p$ and $\lambda_1'\tau$, the condition $p<\frac{1}{2}-\frac{\log \gamma}{2(\gamma-1)}-\gamma_1\tau$ can be satisfied by the photon-counting system under consideration, and thus the optimal holding time can be obtained in Lemma~2.

Then we consider the case of $T\le\tau$, where the KL distance $D(P_0^B||P_1^B)$ can be written as follows
\be\label{eq.KLD3}
D(P_0^B||P_1^B)= N_1\log\frac{1}{1-P_1}+N_1
\log (1-P_0)+N_0P_0\left[\log\frac{P_0}{P_1}+\log(1-P_1)\right]-N_0P_0\log(1-P_0).
\ee

We first discuss the term $N_1\log\frac{1}{1-P_1}$. Considering sufficiently small $p$ such that $p<<\lambda_1' T$, we have $P_1\approx2\tau'\hat N_1$ and $N_1\approx \frac{1}{2\tau'}$. Then the first term of KL distance $D(p_0^B||p_1^B)$ given in (\ref{eq.KLD3}) can be approximated by
\be
N_1\log\frac{1}{1-P_1}\approx 2\tau\hat N_1\log\frac{1}{1-2\tau'\hat N_1}.
\ee

We first have the following lemma that provides a sufficient condition on the negative property of the derivative of $N_1\log\frac{1}{1-P_1}$ with respect to $\tau$.

\begin{lemma}
If probability $p\le 1-e^{-\lambda_1'^3T^3}$, we have that
$N_1\log\frac{1}{1-P_1}$ decreases strictly with respect to $\tau$ for $\tau\ge T$.

\begin{proof}
Please refer to Appendix.G.
\end{proof}
\end{lemma}

For the term $N_1\log(1-P_0)$, note that $N_1=\frac{1}{2\tau'}$ decreases with respect to $\tau$, and
\be
\frac{\partial P_0}{\partial \tau}=2\hat N_0\left[1-\lambda_0\tau'-\frac{p(1-\lambda_0\delta)}{p+\lambda_0T}\right]\ge0,
\ee
if $p\le\frac{1-\lambda_0\tau'}{\alpha+1/2}$.
Thus the term $N_1\log(1-P_0)$ decreases with respect to $\tau$ if $p\le\frac{1-\lambda_0\tau'}{\alpha+1/2}$.

We consider the second term $N_0P_0\left[\log\frac{P_0}{P_1}+\log(1-P_1)\right]$, denoted as $R(\tau)$. Its derivative with respect to $\tau$ is given as follows,
\be
\frac{\partial R(\tau)}{\partial \tau}=\hat N_0\frac{\partial r(\tau)}{\partial \tau}+\frac{\partial \hat N_0}{\partial \tau}r(\tau),
\ee
where $r(\tau)\dff\log\frac{P_0}{P_1}+\log(1-P_1)$.
For the function $r(\tau)$, we have the following lemma.
\begin{lemma}
If $p\le1-\frac{2(\alpha-1)}{2\alpha+1}e^{\lambda_0'(\tau+T)}$, then function $r(\tau)$ strictly decreases with respect to $\tau$.

\begin{proof}
Please refer to Appendix.H.
\end{proof}

Finally, according to the above results on the terms of $D(P_0^B||P_1^B)$, assuming that the condition $p\le\min\left\{1-e^{-\lambda_1'^3T^3},1-\frac{2(\alpha-1)}{2\alpha+1}e^{\lambda_0'(\tau+T)}\right\}$ is satisfied, we have the following,
\be\label{eq.KLD_difference}
D_{01}(\tau)-D_{01}(T)&\le&\max_{\tau} \frac{\partial \hat N_0}{\partial \tau}\left[\log\frac{P_0}{P_1}+\log(1-P_1)\right](\tau-T)+{\hat N_0 P_0} \nonumber \\
&=&\max_{\tau} \left\{P_0-\lambda_0'(\tau-T)\left[\log\frac{P_0}{P_1}+\log(1-P_1)\right]\right\}\hat N_0.
\ee
According to (\ref{eq.opt_tau_ineq1}), we have $\frac{\hat N_0}{\hat N_1}\ge\frac{\lambda_0'}{\lambda_1'}$ and $\frac{N_1}{N_0}\ge\frac{\tau_0'}{\tau'}\ge\frac{3}{2\alpha+3}$, which provides a lower bound on $\frac{P_0}{P_1}$, given by $\frac{P_0}{P_1}=\frac{\hat N_0}{\hat N_1}\frac{N_1}{N_0}\ge\frac{3\lambda_0'}{(2\alpha+3)\lambda_1'}$. Then an upper bound on $D_{01}(\tau)-D_{01}(T)$ is given follows,
\be\label{eq.KLD_difference2}
D_{01}(\tau)-D_{01}(T)\le\max_\tau\left\{P_0+\lambda_0'\tau\log\frac{1}{\lambda_0'}+\lambda_0'\tau\left[\log\frac{(2\alpha+3)\lambda_1'}{3}+\log(1-P_1)\right]\right\}\hat N_0.
\ee
In optical scattering communication, we consider sufficiently small $\lambda_0$ and thermal noise variance such that $\lambda_0<<1$ and $\frac{p}{T}<<1$. Then we have that $\hat N_0<\frac{1-(1-\lambda_0'T)(1-p)}{T}<\lambda_0'+\frac{p}{T}<<1$ and $P_0<2\tau\hat N_1<<1$.
It is seen that for not large $\alpha$ and $\lambda_1$ (large $\lambda_1$ would incur channel capacity loss, because we can divided each slot into a shorter one to realize a higher transmission rate), the term $\lambda_0'\tau\left[\log\frac{(2\alpha+3)\lambda_1'}{3}+\log(1-P_1)\right]$ can be regarded as the same order of $\lambda_0'\tau$. Hence, it could be observed that the expression of Equation (\ref{eq.KLD_difference2}) is of the order lower than $-\bar N_0^2\log\bar N_0$, where $\bar N_0\dff\lambda_0'+\frac{p}{T}$. Since $\bar N_0<<1$, the upper bound given in (\ref{eq.KLD_difference2}) is also small and can be neglected. Based on the experimental measurements, the detail numerical analysis of this upper bound will be further presented in Section \ref{section_V_C} .

Hence, we choose $\tau^*=T$ as the holding time, and the performance loss may become negligible.
\end{lemma}
\subsubsection{Optimization on $\xi$}
Recall that we select the value $\tau^*=T$ as the sub-optimal but satisfactory holding time, which is independent of the decision threshold $\xi$. Then we have that $\alpha=1$, $\delta=0$ when $\tau=T$, thus Equations (\ref{eq.KLD1}) and (\ref{eq.KLD2}) are identical.

We aim to maximize $D(P_0^B||P_1^B)$. Due to the complicated form of the KL distance with respect to $\xi$, we adopt numerical optimization solutions, such as Genetic Algorithm to seek the optimal decision threshold. Note that the KL distance is a bounded deterministic function with respect to $p$ and $q$ when $\lambda_1$ and $\lambda_0$ are given. Considering small $\sigma_0$ and $\sigma$, there exists a wide region of $\xi$ where the parameters $p$ and $q$ vary slightly, which implies that the KL distance also varies slightly. In the numerical results, we will show that the BER curve with respect to $\xi$ has a wide and flat region, where the performance loss of optimal threshold selection deviation is negligible.

\subsection{Discussions of the Conditions on $p$ and Upper Bound in (\ref{eq.KLD_difference2})}\label{section_V_C}

\subsubsection{Conditions on $p$}
In the previous subsection, we provides three conditions on probability $p$, which can be summarized as follows,
\be\label{eq.diss_p_1}
p\le\min\left\{1-e^{-\lambda_1'^3T^3},\frac{1-\lambda_0\tau'}{\alpha+1/2},1-\frac{2(\alpha-1)}{2\alpha+1}e^{\lambda_0'(\tau+T)}\right\}.
\ee

In the UV optical wireless communication, the background radiation can be extremely small, where we assume $\lambda_0<<1$ and $\frac{\lambda_1}{\lambda_0}>>1$.

Considering the term $1-e^{-\lambda_1'^3T^3}$. In the receiver side, we assume the sampling resource is well utilized, which implies $\lambda_1'T$ could not be extremely small like $\lambda_0'T$. Thus, we assume $\lambda_1'T>0.02$ by referring to our experiment setting and channel characterization experiment \cite{wang2017demonstration}. Then we have $1-e^{-\lambda_1'^3T^3}>8\times 10^{-6}$.

For the second term, we assume extremely small $\lambda_0'(\tau+T)$ $(<0.01)$ and not large $\alpha$ $(\le10)$. Then it can be seen that $\frac{1-\lambda_0\tau'}{\alpha+1/2}>0.0943$.

As for the last term in the right side of (\ref{eq.diss_p_1}), according to the above assumption on $\alpha$ and $\lambda_0\tau'$, it can be observed that $1-\frac{2(\alpha-1)}{2\alpha+1}e^{\lambda_0'(\tau+T)}>0.144$.

For the photon counting system, the variance thermal is found to be significantly less than one $(<0.02^2)$, then it could be seen that for $\xi>0.09$, we have $p<Q(\frac{\xi}{\sigma_0})=3.4\times10^{-6}<8\times 10^{-6}$. Thus, the conditions on $p$ could be satisfied when the decision threshold $\xi$ is not too small.

\subsubsection{Discussion on Upper Bound in (\ref{eq.KLD_difference2})}
Considering that for sufficiently weak background radiation, i.e. $\lambda_0<0.1$, $\lambda_1=20$ is large enough to achieve reliable information transmission, and we may shorten each time slot for a larger $\lambda_1$ to increase the communication rate. Meanwhile, assuming $\lambda_1\tau'<<1$ such that $P_1=2\tau'\hat N_1\le2\tau'\lambda_1\le0.5$, and $\alpha\le10$, we have $\log\frac{(2\alpha+3)\lambda_1'}{3}-\log(1-P_1)<5.7258$. Note that we have $\lambda_0'\tau<0.01$ for sufficiently weak background radiation, $T>\frac{0.02}{20}=0.001$ for well utilization of sampling resource, and $p<8\times 10^{-6}$ for satisfying the conditions $p\le1-e^{-\lambda_1'^3T^3}$. It can be seen that $D_{01}(\tau)-D_{01}(T)<0.0102<<1$, which implies that the photon counting system only incur negligible performance loss compared with the optimal one when we select $\tau^*=T$.

\subsection{Signal Detection}
Since we select $\tau^*=T$ as the sub-optimal holding time in the photon-counting system, it can be seen that $N_1=N_0=\frac{1}{3T}$. Thus the two likelihood functions are given as follows,
\be
\mathbb P(n|\lambda_1,\tau=T)&\approx&B(\frac{1}{3T},3T\hat N_1);\nonumber \\
\mathbb P(n|\lambda_0,\tau=T)&\approx&B(\frac{1}{3T},3T\hat N_0).
\ee
Based on the above likelihood functions, we can obtain the following detection threshold,
\be
\hat n_{th}=\left\lfloor \frac{1}{3T}\frac{\log\frac{1-3T\hat N_1}{1-3T \hat N_0}}{\log\frac{\hat N_1}{\hat N_0}+\log\frac{1-3T\hat N_1}{1-3T \hat N_0}}\right\rfloor.
\ee
The error probabilities are then given by
\be\label{eq.error_prob2}
p_e=\frac{p_{10}+p_{01}}{2}=\frac{\sum_{n=0}^{\hat n_{th}}\mathbb P(n|\lambda_1,\tau)+\sum_{\hat n_{th}+1}^{\infty}\mathbb P(n|\lambda_0,\tau)}{2}.
\ee

\section{Experimental and Numerical Simulation Results}
\begin{figure}[t!]
\centering
\includegraphics[width = 1\columnwidth]{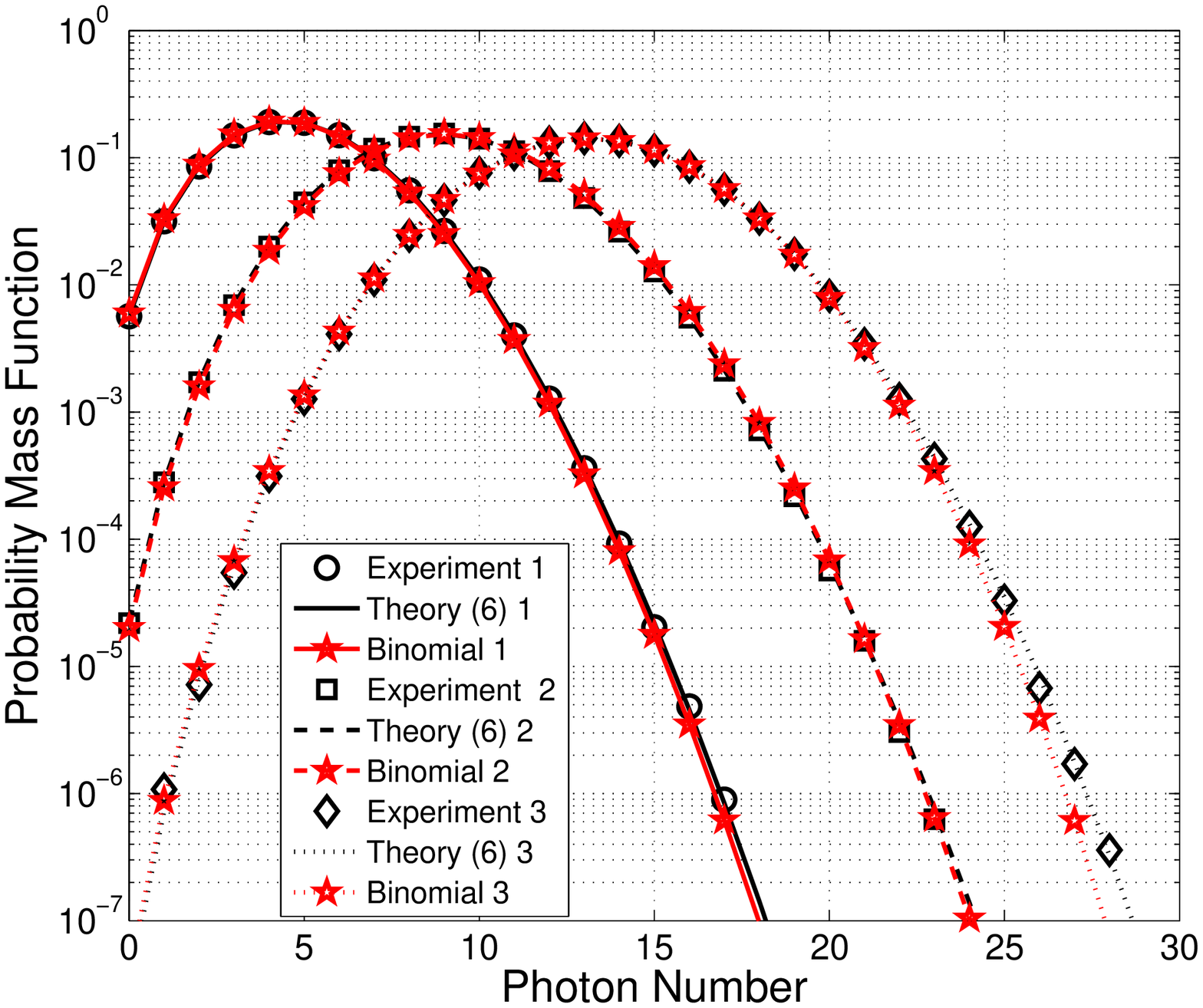}
\caption{The PMF of recorded pulses per microsecond: comparison between experimental, the distribution in Equation (6), and the binomial distribution fitting.}
\label{fig.exp_fit}
\end{figure}

\begin{figure}[t!]
\centering
\includegraphics[width = 1\columnwidth]{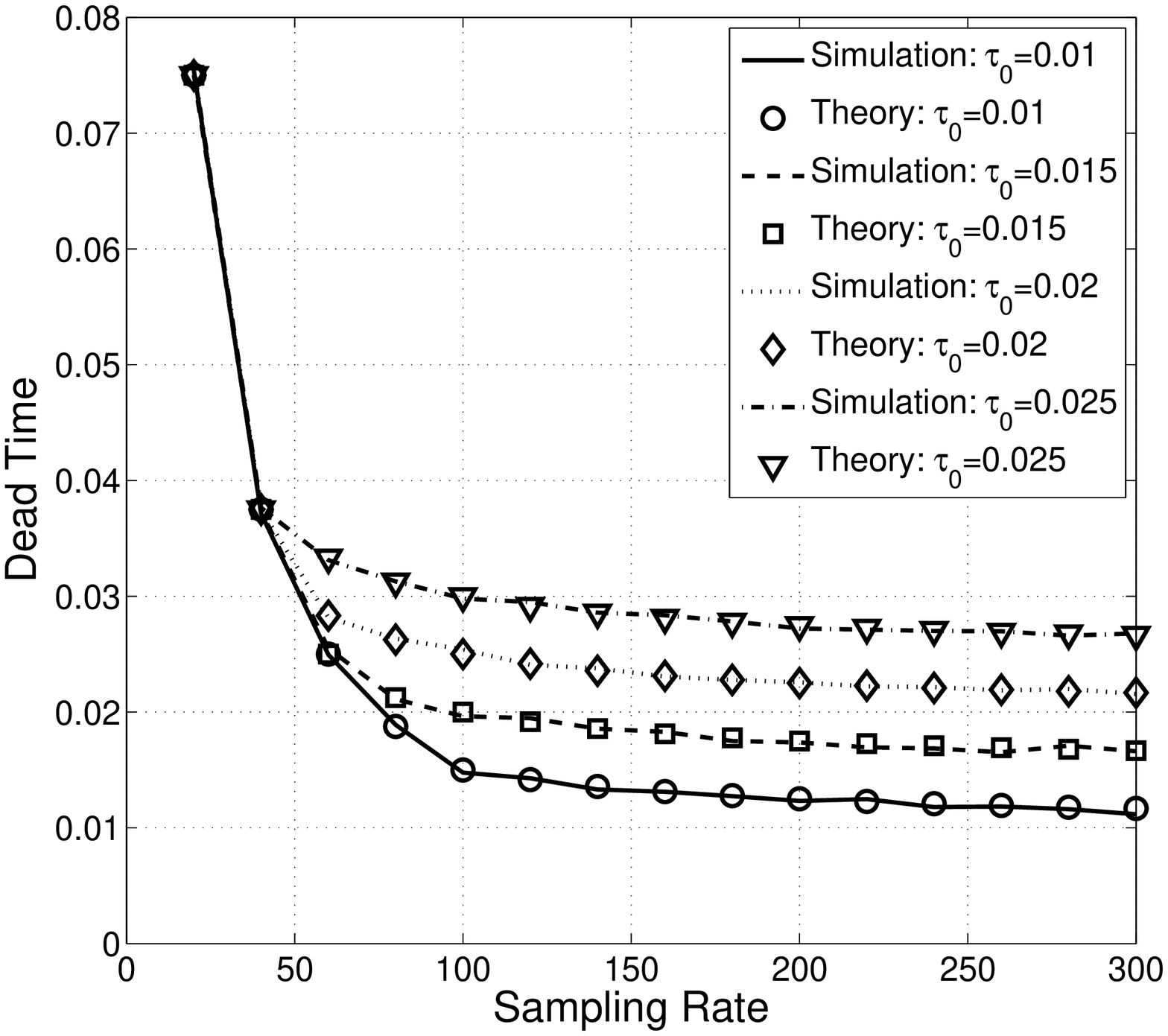}
\caption{Equivalent dead time with respect to the sampling rate for different holding times.}
\label{fig.fit_nonoise1}
\end{figure}
\begin{figure}[t!]
\centering
\includegraphics[width = 1\columnwidth]{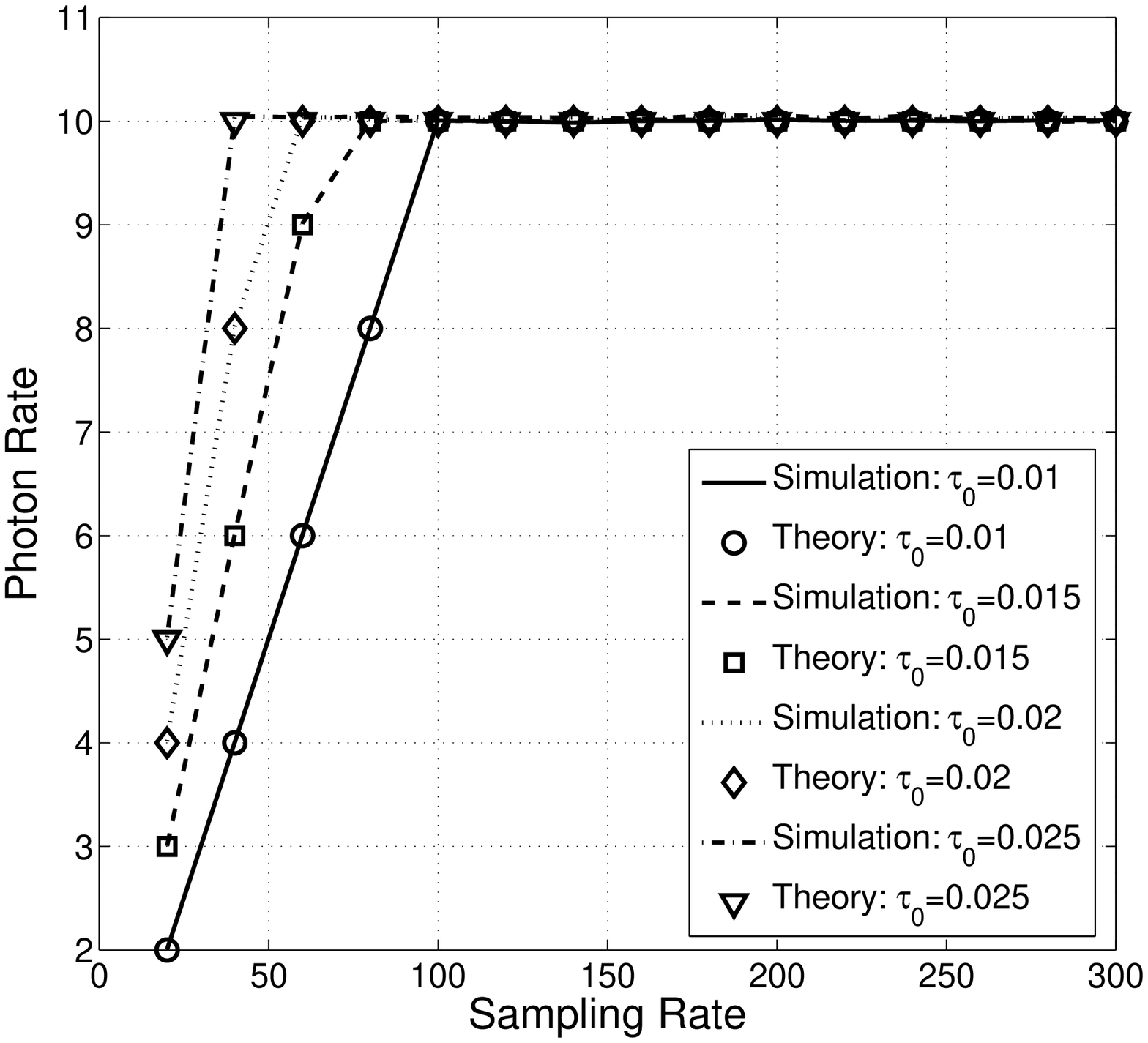}
\caption{Equivalent photon arrival rate with respect to the sampling rate for different holding times. }
\label{fig.fit_nonoise2}
\end{figure}

\begin{figure}[t!]
\centering
\includegraphics[width = 1\columnwidth]{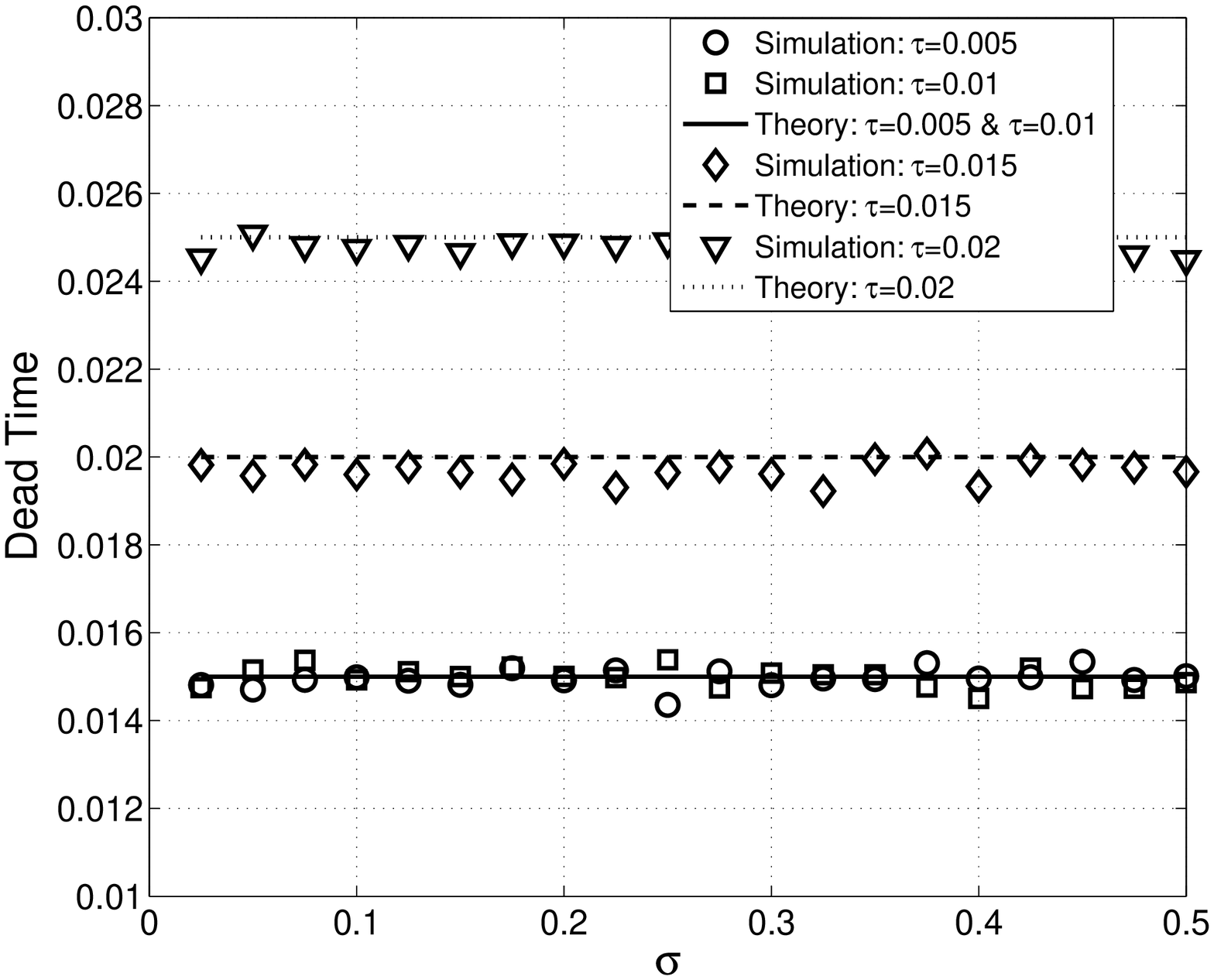}
\caption{Equivalent dead time with respect to the shot noise variances for different holding times. }
\label{fig.fit_noise1}
\end{figure}
\begin{figure}[t!]
\centering
\includegraphics[width = 1\columnwidth]{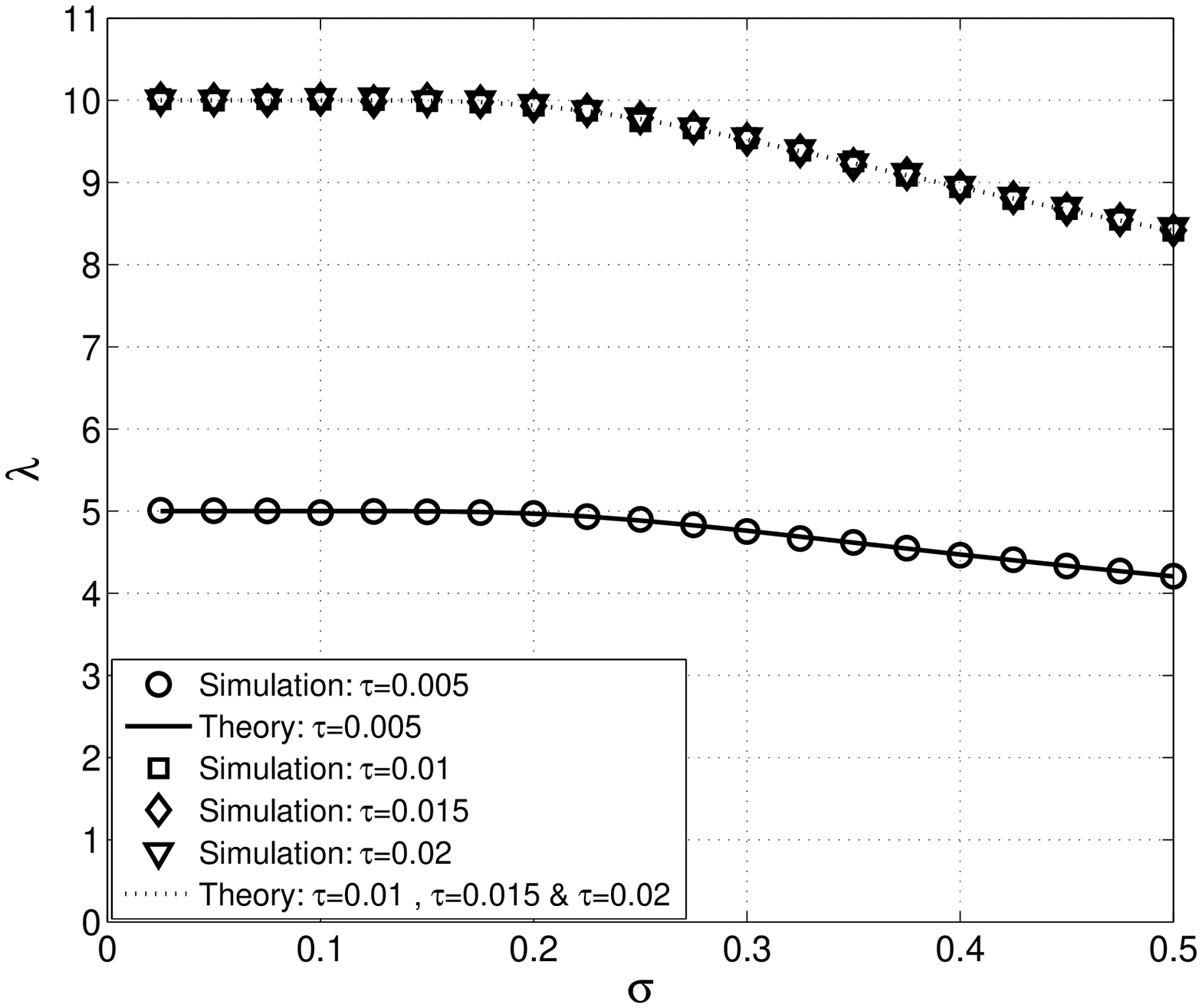}
\caption{Equivalent photon arrival rate with respect to the shot noise variances for different holding times. }
\label{fig.fit_noise2}
\end{figure}

\begin{figure}[t!]
\centering
\includegraphics[width = 1\columnwidth]{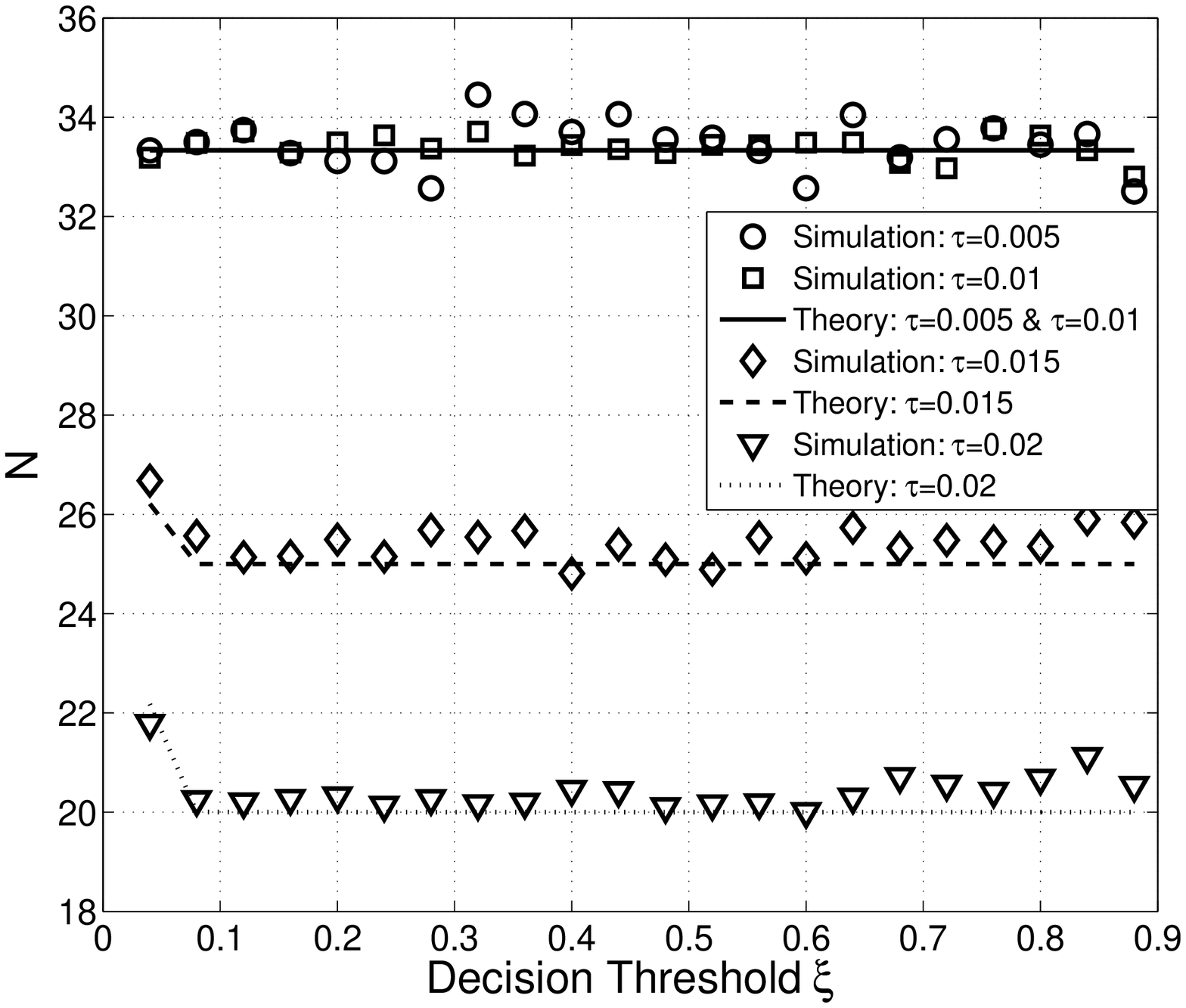}
\caption{The binomial distribution parameter $N$ with respect to the decision threshold for different holding times. }
\label{fig.fit_noise3}
\end{figure}
\begin{figure}[t!]
\centering
\includegraphics[width = 1\columnwidth]{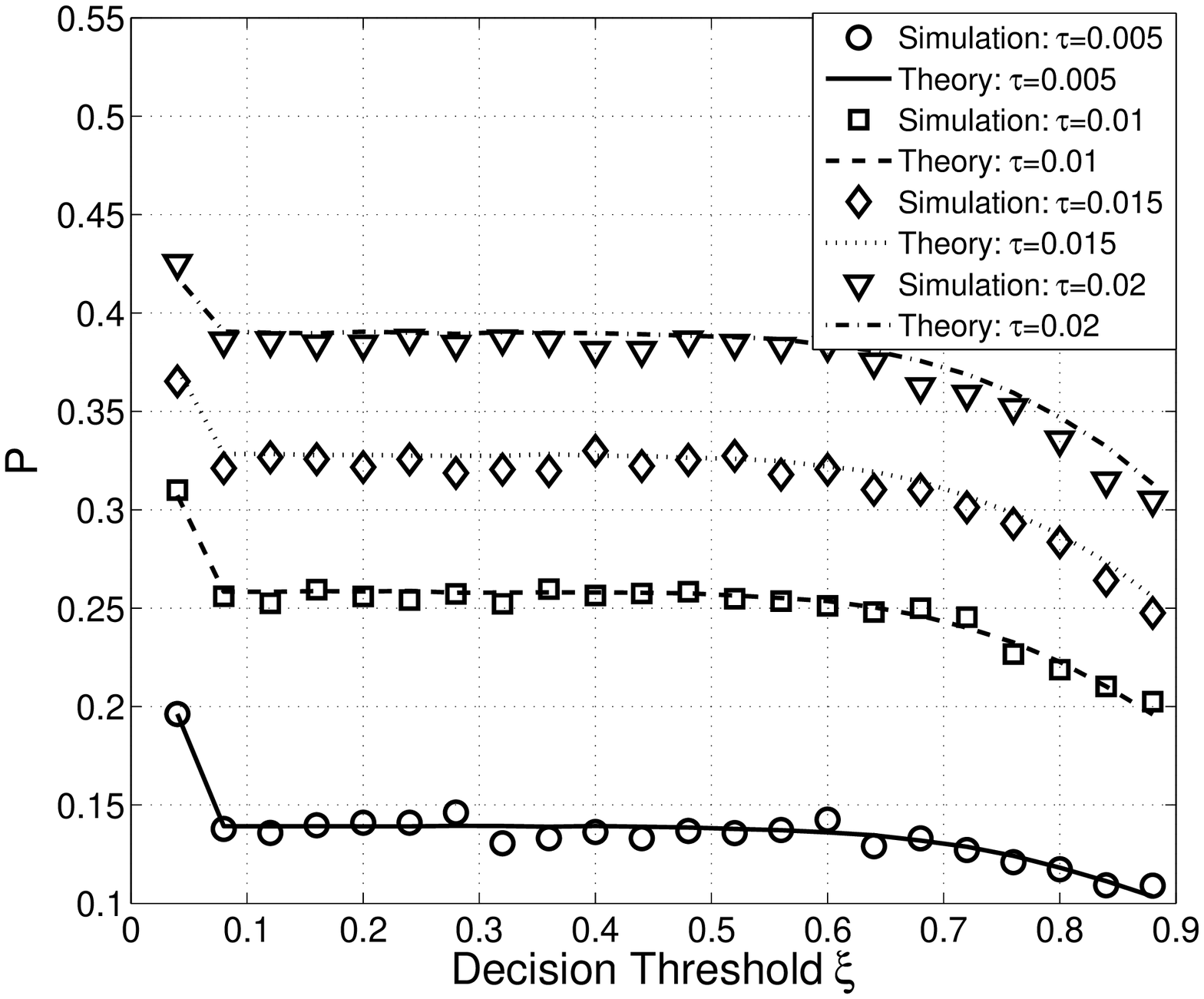}
\caption{The binomial distribution parameter $P$ with respect to the decision threshold for different holding times. }
\label{fig.fit_noise4}
\end{figure}
\begin{figure}[t!]
\centering
\includegraphics[width = 1\columnwidth]{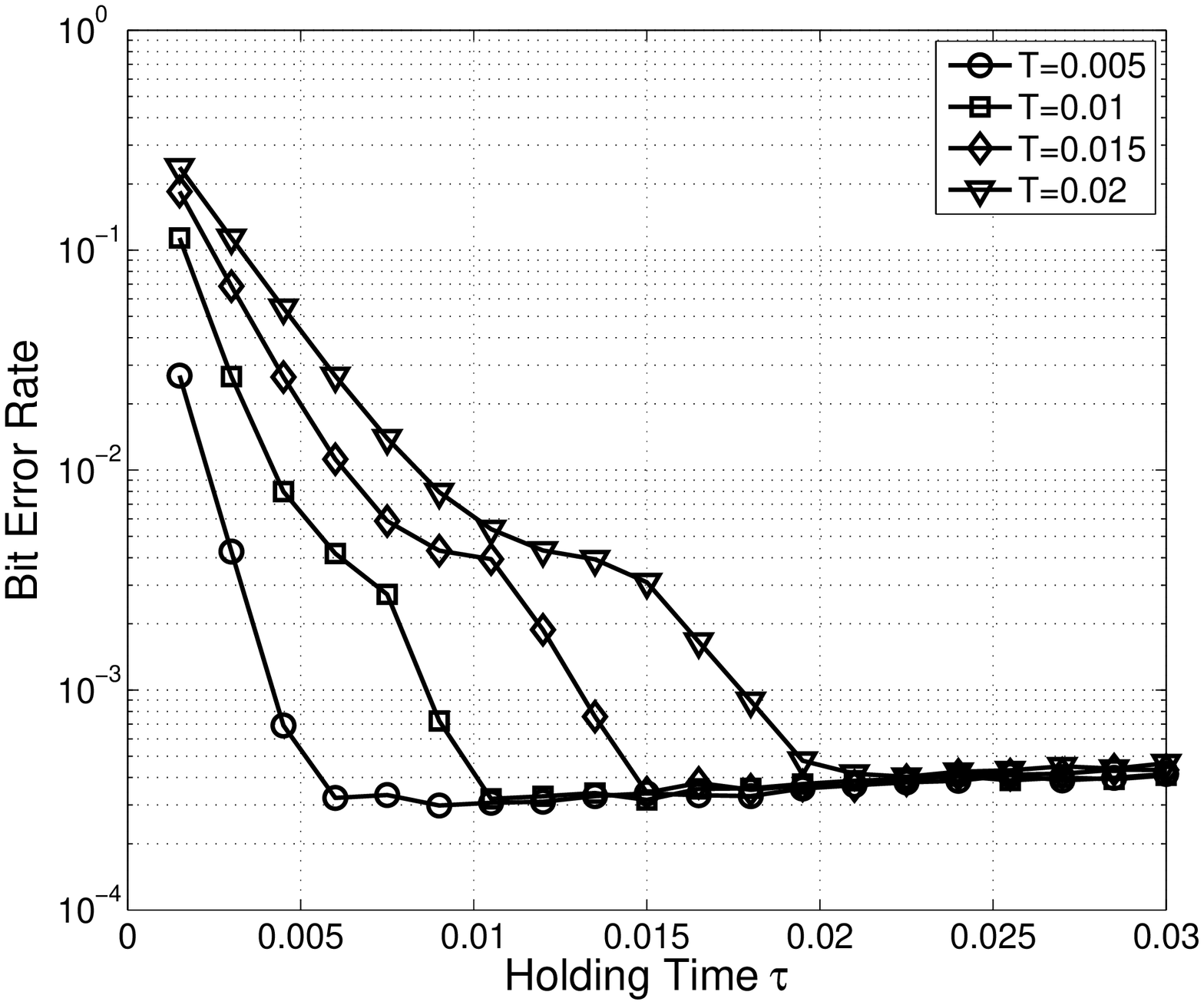}
\caption{BER of the photon counting system with respect to the holding time for different sampling periods. }
\label{fig.ber_sampling1}
\end{figure}

\begin{figure}[t!]
\centering
\includegraphics[width = 1\columnwidth]{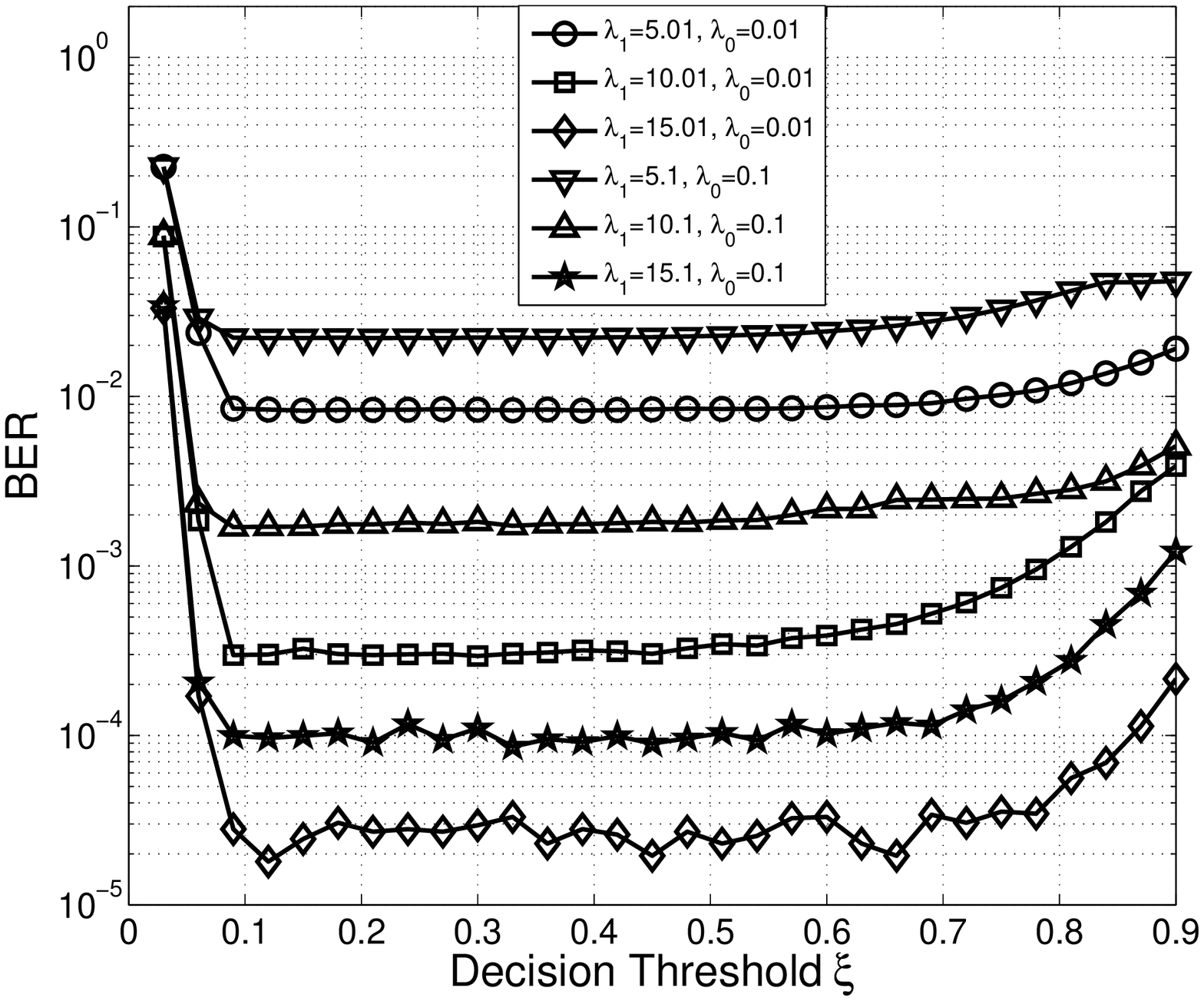}
\caption{BER of the photon counting system with respect to the decision threshold for different signal and background radiation intensities . }
\label{fig.ber_th1}
\end{figure}

\begin{figure}[t!]
\centering
\includegraphics[width = 1\columnwidth]{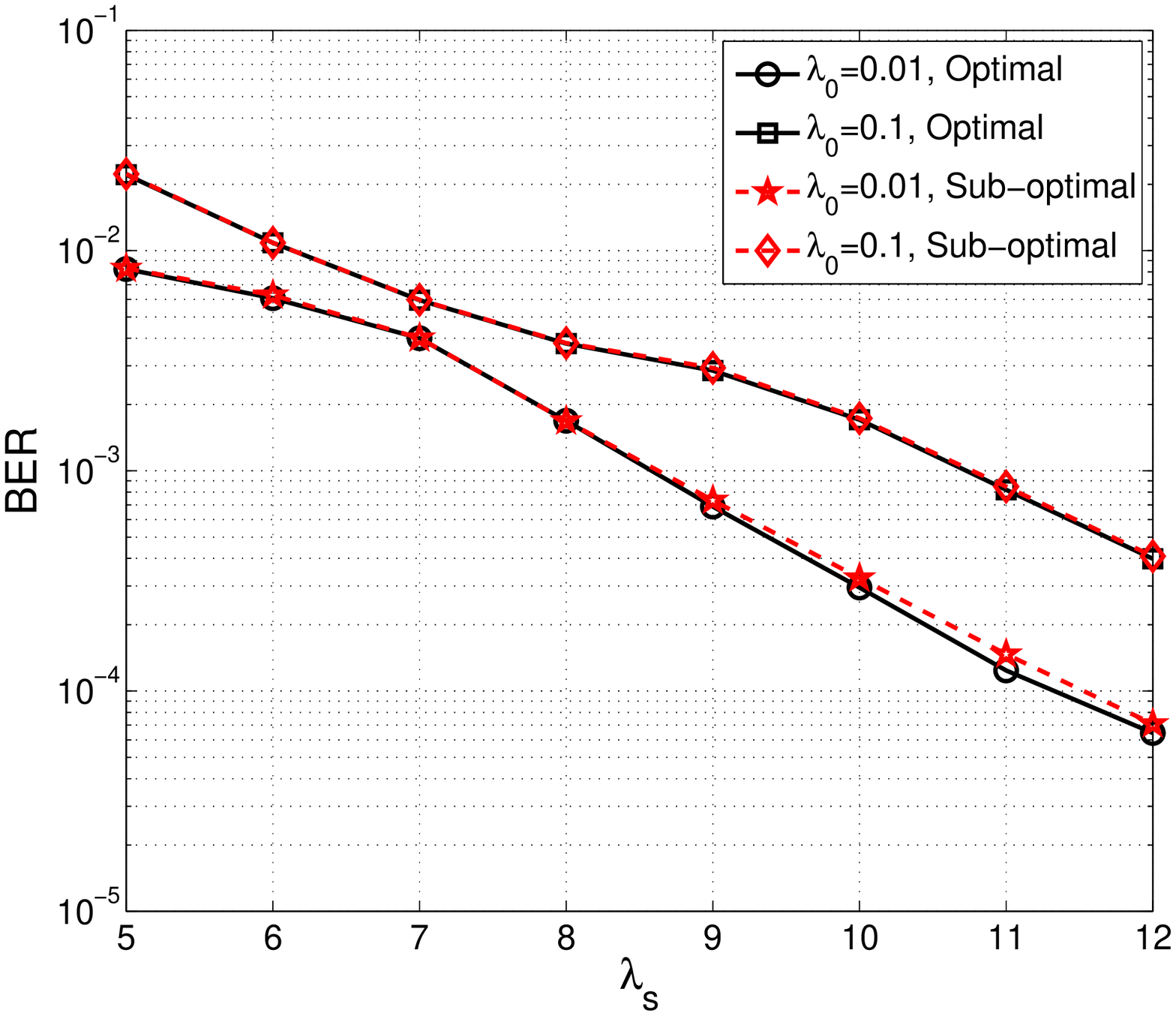}
\caption{The BER of the photon counting system with respect to the photon arrival rate $\lambda_s$ for sub-optimal selection rule and the optimal counterpart. }
\label{fig.ber_compare}
\end{figure}

We first present the experimental results. In the transmitter side, the intensity of transmitted light remains constant during the photon-counting process, while three experiments with different intensities are conducted. In the receiver side, a PMT, pulse holding circuits, an ADC, and a post-processing FPGA are adopted to realize the photon-counting process. In the three different experiments, the ADC sampling rate is set to be $100$MHz, and the decision threshold is set to be a low value due to the small thermal noise.

Figure~\ref{fig.exp_fit} shows the PMF of recorded pulses per microsecond, which contains the experimental results, the fitting results according to Equation (\ref{eq.dis_subpoisson}) and binomial distribution. The fitting parameters are obtained based on matching the first-order and second-order moments.  It can be seen that Equation (\ref{eq.dis_subpoisson}) and binomial distribution both fit well even under finite sampling rate and electrical noise. The binomial distribution can serve as a good model to describe the number of recorded photoelectrons.

Then we provide simulation results to verify our proposed approximation results in Section.II.  Assume the mean number of photoelectrons $\lambda=10$, and no electrical noise. Figure~\ref{fig.fit_nonoise1} and Figure~\ref{fig.fit_nonoise2} show the estimated equivalent dead time and photon arrival rate with respect to sampling rate for different holding times,  where the results from the both theoretical analysis and simulation are provided. It is seen that the two types of results match well, which validates Corollary~$1$ and~$2$ in Section.II.

Moreover, we consider the  photon-counting system with shot noise and finite sampling rate. Assuming the mean number of photoelectrons $\lambda=10$, and $100$ samples per symbol duration. Figures~\ref{fig.fit_noise1} and~\ref{fig.fit_noise2} show the equivalent dead time $\tau'$ and photon arrival rate $\lambda'$ compared with the ideal model for different shot noise variances, respectively, based on both theoretical approximation and simulations. It can be seen that the equivalent dead time and photon arrival rate obtained from simulations match well with the theoretical results given in Corollary~$3$ and~$4$. We also consider thermal noise, where the shot and thermal variances are $0.2$ and $0.02$, respectively. Figures~\ref{fig.fit_noise3} and~\ref{fig.fit_noise4} compare the binomial distribution parameters $N$ and $P$ obtained from numerical simulation and theoretical results of Theorem~3 for different decision thresholds, respectively. It can also be seen that the binomial approximation with parameters given in Theorem~3 can well characterize the practical photon-counting system with both thermal and shot noise under finite sampling rate.

We adopt Monte Carlo method to obtain the bit error rate of the photon counting system, where the shot and thermal variances are set to be $0.2$ and $0.02$, respectively.
Figure~\ref{fig.ber_sampling1} shows the simulation results of bit error rate performance for different holding times $\tau$ and sampling periods $T$, where decision threshold $\xi$ is set to be $0.3$. It can be seen that the photon counting system performs well when $\tau=T$ for fixed $T$, the BER performance is close to the optimal one. Figure~\ref{fig.ber_th1} shows the simulated BER for different decision thresholds $\xi$, where $\tau$ is set to be $0.01$ according to the holding time selection rule. It can be observed that there exists a wide and flat region in each curve, where the optimal threshold locates in such flat region while a slight change of threshold selection may only incur negligible performance loss. Finally, we compare the performance of the proposed sub-optimal holding time and decision threshold selection rule and the optimal counterpart in Figure~\ref{fig.ber_compare}. It can be seen that the proposed sub-optimal selection rule shows negligible performance loss compared with the optimal one.

\section{Conclusion}
We have analyzed the architecture of PMT-based photon-counting receiver with finite holding time and sampling rate, and showed that the dead time effect can lead to a sub-Poisson characteristics. We have studied the first-order and second-order moments on the sub-Poisson for the number of detected photoelectrons under finite sampling rate and electrical noise. Moreover, we have proposed a binomial distribution approximation on such sub-Poisson distribution and provided a tractable holding time and decision threshold selection rule based on maximizing the minimal KL distance. Experimental results showed that the proposed sub-Poisson model and the binomial approximation can well characterize practical photon-counting system. Besides, numerical results can well characterize the equivalent arrival rate under finite-rate sampling and the associated binomial parameters $P$ and $N$ under electrical noises. Simulations results also shown that the performance of the proposed holding time and decision threshold selection rules is close to that of the optimal counterpart.

\section{Appendix}
\subsection{Proof of Theorem~1}
Note that $\mathbb P[n[k]=1]=e^{-\lambda \tau}(1-e^{-\lambda \tau})$, we have the following
\be\label{eq.append_A_1}
\mathbb E[n_s]&=&\mathbb E\left[\sum_{k=0}^{N-1}n[k]\right]= \sum_{k=0}^{N-1}\mathbb E[n[k]] \nonumber\\
&=&\frac{e^{-\lambda \tau}(1-e^{-\lambda \tau})}{T}.
\ee

We next consider the variance of $n_s$. We have the following second moment of $n_s$,
\be\label{eq.append_S_2}
\mathbb E[n_s^2]&=&\mathbb E\left[\left(\sum_{k=0}^{N-1}n[k]\right)^2\right] \nonumber\\
&=&\mathbb E \left[\sum_{k=0}^{N-1}n[k]^2+\sum_{k\neq l}n[k]n[l]\right].
\ee
Since $n[k]$ can only take the value of $0$ or $1$, we have that $\mathbb E[n[k]^2]=\mathbb E[n[k]]$. Thus we have that
\be\label{eq.append_B_3}
\mathbb E \left[\sum_{k=0}^{N-1}n[k]^2\right]=\mathbb E[n_s]=\frac{e^{-\lambda \tau}(1-e^{-\lambda T})}{T}.
\ee

Consider the case of $|k-l|>1$, where the photoelectron detected in interval $[kT,(k+1)T]$ has no impact on the counting in interval $[lT,(l+1)T]$. Thus $n[k]$ and $n[l]$ are statistically independent, which shows
\be
\mathbb E[n[k]n[l]]&=&\mathbb E[n[k]]^2\nonumber \\
&=&\left[e^{-\lambda\tau}(1-e^{-\lambda\tau})\right]^2, \ \ \ for \ |k-l|>1.
\ee
Consider the case of $|k-l|=1$. Since the sample at time $(k+1)T$ must be larger than the threshold if one pulse is detected in interval $[kT,(k+1)T]$, we have $\mathbb P[n[l]=0|n[k]=1]=1$, and thus
\be
\mathbb E[n[k]n[l]]=0, \ \ \ for \ |k-l|=1.
\ee
Based in the above two cases, we have
\be
\mathbb E\left[\sum_{k\neq l}n[k]n[l]\right]&=&\sum_{|k-l|\ge2}\mathbb E[n[k]n[l]]
\nonumber \\
&=&(\frac{1}{T}-1)(\frac{1}{T}-2)\left[e^{-\lambda\tau}(1-e^{-\lambda\tau})\right]^2,\nonumber \\
\ee
and then
\be
\mathbb E[n_s^2]=\mathbb E[n_s]+(1-3T+2T^2)\mathbb E[n_s]^2.
\ee

\subsection{Proof of Theorem~2}
Note that $\mathbb P[n[k]=1]=e^{-\lambda \tau}(1-e^{-\lambda T})$, we have the following
\be\label{eq.append_B_1}
\mathbb E[n_s]&=&\mathbb E\left[\sum_{k=0}^{N-1}n[k]\right]= \sum_{k=0}^{N-1}\mathbb E[n[k]] \nonumber\\
&=&\frac{e^{-\lambda \tau}(1-e^{-\lambda T})}{T}.
\ee

We next consider the variance of $n_s$. The second moment of $n_s$ is given as follows
\be\label{eq.append_B_2}
\mathbb E[n_s^2]&=&\mathbb E\left[\left(\sum_{k=0}^{N-1}n[k]\right)^2\right] \nonumber\\
&=&\mathbb E \left[\sum_{k=0}^{N-1}n[k]^2+\sum_{k\neq l}n[k]n[l]\right].
\ee
Similar to the case of $T>\tau$, we have $\mathbb E \left[\sum_{k=0}^{N-1}n[k]^2\right]=\mathbb E[n_s]$.

Since the sampling period $T$ is less than or equal to the dead time $\tau$, let $\tau=\alpha T+\sigma$, where $\alpha$ is a positive integer and $0\le\sigma<T$. Due to the dead time effect, if one pulse is detected in interval $[kT,(k+1)T]$, the samples at time $(k+1)T, (k+2)T, \dots, (k+\alpha)T$ must be larger than the threshold, which implies that we cannot detect any pulse in the sampling intervals from $[(k+1)T,(k+2)T]$ to $[(k+\alpha)T,(k+\alpha+1)T]$. Thus we have
\be
\mathbb E[n[k]n[l]]=0, \ \ \  for \ |k-l|\le\alpha.
\ee
If $|k-l|\ge\alpha+2$, the rising edge detection in interval $[kT,(k+1)T]$ has no impact on that in $[lT,(l+1)T]$, and thus the number of detected pulse $n[k]$ and $n[l]$ are statistically independent. Then we have
\be
\mathbb E[n[k]n[l]]&=&\mathbb E[n[k]]^2 \nonumber \\
&=&\left[e^{-\lambda \tau}(1-e^{-\lambda T})\right]^2, \ \  for \ |k-l|\ge\alpha+2.
\ee
The last situation is $|k-l|=\alpha+1$, which needs to be more delicately analyzed. Assuming the number of detected pulse in interval $[kT,(k+1)T]$ is one, the necessary condition of one pulse detected in $[lT,(l+1)T]$ is the sample at time $lT$ smaller than the threshold, or no photon arrival in time region $[lT-\tau,lT]$. Note that since $lT-\tau=(k+\alpha+1)T-\alpha T-\sigma=(k+1)T-\sigma\le(k+1)T$, the event that $n[k]=n[l]=1$ is equivalent to the following two events: no arrival in $[kT-\tau,kT]\cup[lT-\tau,lT]$ and at least one arrival in $[kT,(k+1)T-\sigma]$ and $[lT,(l+1)T]$. The probability of such an event is given by
\be
\mathbb P[n[k]=1,n[l]=1]=e^{-2\lambda \tau}\left(1-e^{-\lambda (T-\sigma)}\right)\left(1-e^{-\lambda T}\right).
\ee
Considering all the above three cases, we have that
\be
\mathbb E\left[\sum_{k\neq l}n[k]n[l]\right]&=&\sum_{|k-l|\ge\alpha+2}\left[e^{-\lambda \tau}(1-e^{-\lambda T})\right]^2+\sum_{|k-l|=\alpha+1}e^{-2\lambda \tau}\left(1-e^{-\lambda (T-\sigma)}\right)\left(1-e^{-\lambda T}\right) \nonumber \\
&=&(1/T-\alpha-1)(1/T-\alpha-2)\left[e^{-\lambda \tau}(1-e^{-\lambda T})\right]^2 \nonumber \\
&&+2(\frac{1}{T}-\alpha-1)e^{-2\lambda \tau}\left(1-e^{-\lambda (T-\sigma)}\right)\left(1-e^{-\lambda T}\right).
\ee
Substituting the above results into (\ref{eq.append_B_3}) can lead to the second moment.

\subsection{Proof of Corollary~3}
Note that the event of $F[kT]<\xi$ and $F[(k+1)T]>\xi$ are independent with each other, we have the following approximated probability of $\mathbb P[n[k]=1]$ based on the approximation forms of (\ref{eq.noisecase1_prob1}) and (\ref{eq.noisecase1_prob2}),
\be
\mathbb P[n[k]=1]&=&\mathbb P[F[kT]<\xi] \mathbb P[F[kT]>\xi] \nonumber \\
&\approx&e^{-\lambda \tau}(1+\lambda \tau q)\left[1-e^{-\lambda \tau}(1+\lambda \tau q)\right].
\ee
Note that $\lambda\tau$ and $q$ are both small, the mean of $n_s$ is approximated by
\be
\mathbb E[n_s]&=&\frac{e^{-\lambda \tau}(1+\lambda \tau q)\left[1-e^{-\lambda \tau}(1+\lambda \tau q)\right]}{T}\nonumber \\
&\approx&\frac{e^{-\lambda\tau}e^{\lambda\tau q}\left(1-e^{-\lambda\tau}e^{\lambda\tau q}\right)}{T}\nonumber \\
&\approx& \frac{(1-q)\lambda\tau}{T}e^{-\frac{(1-q)\tau}{T}\cdot\frac{3T}{2}},
\ee
where the last step follows the same procedure as (\ref{eq.appendA_appro_mean}).

We next consider the variance. The variance of $n_s$ can be approximated as follows
\be
\mathbb D[n_s]=\mathbb E[n_s]+\sum_{k\neq l}\mathbb E[n[k]n[l]]-\mathbb E[n_s]^2.
\ee
Similar to the case without shot or thermal noise, we have $\mathbb E[n[k]n[l]]=0$ for $|k-l|=1$, and $\mathbb E[n[k]n[l]]=E[n_s]^2/T^2$ for $|k-l|\ge 2$. Thus, the variance $\mathbb D[n_s]$ can be approximated by
\be
\mathbb D[n_s]&=&\mathbb E[n_s]+T^2(\frac{1}{T}-1)(\frac{1}{T}-2)\mathbb E[n_s]^2-\mathbb E[n_s]^2 \nonumber \\
&\approx&\mathbb E[n_s]-3T\mathbb E[n_s]^2.
\ee
\subsection{Proof of Corollary~4}
Similar to the case of $T>\tau$, the mean of $n_s$ is approximated by
\be
\mathbb E[n_s]&=&\frac{e^{-\lambda \tau}(1+\lambda \tau q)\left[1-e^{-\lambda T}(1+\lambda T q)\right]}{T}\nonumber \\
&\approx&\frac{e^{-\lambda\tau}e^{\lambda\tau q}\left(1-e^{-\lambda T}e^{\lambda T q}\right)}{T}\nonumber \\
&\approx& (1-q)\lambda e^{-(1-q)\lambda(\tau+\frac{T}{2})},
\ee
where the last step follows the same procedure as (\ref{eq.appendB_approx_mean}).

Then we consider variance of $n_s$. We find the parameter $\alpha$ $0 \leq \delta < T$ such that $\tau=\alpha T+\delta$, the variance can be expressed as follows
\be
\mathbb D[n_s]=\mathbb E \left[\sum_{k=0}^{N-1}n[k]^2+\sum_{|k-l|\ge\alpha+1}n[k]n[l]\right]-\mathbb E[n_s]^2,
\ee
where
\be
\mathbb E[\sum_{k=0}^{N-1}n[k]^2]=\mathbb E[n_s],
\ee
and we have the following for $|k-l|\ge\alpha+2$,
\be
\mathbb E\left[\sum_{|k-l|\ge\alpha+2}n[k]n[l]\right]&=&(\frac{1}{T}-\alpha-1)(\frac{1}{T}-\alpha-2)\mathbb E[n[k]]^2. \nonumber \\
&\approx&(1-T(2\alpha+3))\mathbb E[n_s]^2.
\ee
As for $|k-l|=\alpha+1$, note that more than one photon arrives in $[kT-\tau,kT]$, the probability $p(F[kT]<\xi)$ is assumed to be zero. Similar to the analysis for the case without shot noise, the event that $n[k]=n[l]=1$ is equivalent to the events no photon arrival in $[kT-\tau,kT]\cup[lT-\tau,lT]$ and at least one photon arrival in $[kT,(k+1)T-\sigma]$ and $[lT,(l+1)T]$, respectively. Thus we have
\be
\mathbb E[n[k]n[l]]&=&e^{-\lambda \tau}(1+\lambda \tau q)\left[1-e^{-\lambda (T-\delta)}(1+\lambda(T-\delta)q)\right]\nonumber \\
&&\ e^{-\lambda\tau}(1+\lambda \tau q)\left[1-e^{-\lambda T}(1+\lambda Tq)\right] \nonumber \\
&=&\mathbb E[n[k]]^2\frac{1-e^{-\lambda (T-\delta)}(1+\lambda(T-\delta)q)}{1-e^{-\lambda T}(1+\lambda Tq)}\nonumber \\
&\approx& \mathbb E[n[k]]^2\frac{(\alpha+1)T-\tau}{T}.
\ee
Moreover, we have
\be
\hspace{-5mm}\mathbb E\left[\sum_{|k-l|=\alpha+1}n[k]n[l]\right]&=&2(\frac{1}{T}-1)\mathbb E[n[k]n[l]] \nonumber \\
\hspace{-5mm}&\approx& \mathbb E[n_s]^2[2(\alpha+1)T-2\tau].
\ee
Thus similar to (\ref{eq.appendB_approx_var}), the variance of $n_s$ is approximated by
\be
\mathbb D[n_s]&=&\mathbb E[n_s^2]-\mathbb E[n_s^2] \nonumber \\
 &\approx& \mathbb E[n_s]+\mathbb E[n_s]^2[-T(2\alpha+3)+2(\alpha+1)T-2\tau] \nonumber \\
 &=&\mathbb E[n_s]-2(\tau+\frac{T}{2})\mathbb E[n_s]^2.
\ee

\subsection{Proof of Corollary 6}
Based on the probability of one pulse detected in $[kT,(k+1)T]$, the mean of $n_s$ can be directly derived as follows
\be
\mathbb E[n_s]&=&\sum_{k=0}^{N-1}\mathbb E[n[k]]\nonumber \\
&=&\frac{e^{-\lambda'\tau}(1-p)\left[1-e^{-\lambda'T}(1-p)\right]}{T}.
\ee

As for the variance of $n_s$, we consider the parameter $\alpha$ and $\delta$ such that  $\tau=\alpha T+\delta$, where $0\le\delta<T$. And the variance $\mathbb D[n_s]$ has the following form
\be
\mathbb D[n_s]&=&\mathbb E[n_s^2]-\mathbb E[n_s]^2 \nonumber \\
&=&\mathbb E\left[\sum_{k=0}^{N-1}n[k]^2+ \sum_{k\neq l}n[k]n[l]\right]-\mathbb E[n_s]^2.
\ee
Noting that $\mathbb E\left[\sum_{k=0}^{N-1}n[k]^2\right]=\mathbb E[n_s]$, we analyze the term  $\mathbb E \left[\sum_{k\neq l}n[k]n[l]\right]$ in the following.

First it is obviously that $\mathbb P[n[k]n[l]]=0$ if $|k-l|=1$, since the adjacent two rising-edge cannot exist in the pulse detecting system. Similar to the no noise counting system, we have $\mathbb E[n[k]n[l]]=\mathbb E[n[k]]^2$ for $|k-l|\ge\alpha+2$, since the number of detected pulse $n[k]$ and $n[l]$ are statistically independent.

For $|k-l|=\alpha+1$, $n[k]$ and $n[l]$ are no longer statistically independent. For example, if $n[k]n[l]=1$, then there cannot exist any photon arrival in $[(k+1)T-\delta,(k+1)T]$. Recall that the probability $\mathbb P\left[n[k]=1\right]=e^{-\lambda'\tau}(1-p)\left[1-e^{-\lambda'T}(1-p)\right]$, the probability of $\mathbb P[n[k]n[l]=1]$ is given by
\be
\mathbb P[n[k]n[l]=1]
&=&\mathbb P[n[k]=1|n[l]=1]\mathbb P[n[l]=1] \nonumber \\
&=&e^{-\lambda'\tau}(1-p)\left[1-e^{-\lambda'(T-\delta)}(1-p)\right]\nonumber \\
&& e^{-\lambda'\tau}(1-p)\left[1-e^{-\lambda'T}(1-p)\right] \nonumber \\
&=&\mathbb E[n[k]]^2\frac{1-e^{-\lambda'(T-\delta)}(1-p)}{1-e^{-\lambda'T}(1-p)}.
\ee

In contrast to the case of noiseless counting system, the probability of $n[k]n[l]=1$ is not zero when $1<|k-l|\le\alpha$. Since the thermal noise in different samples are statistically independent, events $n[k]=1$ and $n[l]=1$ may occur in case of no photon arrival in $[kT,lT]$. Therefore, the probability $\mathbb P[n[k]n[l]=1]$ is given by
\be
\mathbb P[n[k]n[l]=1]
&=&\mathbb P[n[k]=1|n[l]=1]\mathbb P[n[l]=1] \nonumber \\
&=&e^{-\lambda'\tau}(1-p)e^{-\lambda'T|l-k|}p(1-p)\left[1-e^{-\lambda'T}(1-p)\right] \nonumber \\
&=&\mathbb E[n[k]]p(1-p)e^{-\lambda'T|l-k|}.
\ee
Thus, the variance $\mathbb D[n_s]$ can be obtained as follows
\be\label{eq.appendix5_var1}
\mathbb D[n_s]&=&\mathbb E[n_s]\big[1+2\sum_{s=2}^\alpha(1-sT)p(1-p)e^{-\lambda'sT}\big]\nonumber \\
&&\ +\mathbb E[n_s]^2\big[(1-(\alpha+1)T)(1-(\alpha+2)T)-1 \nonumber \\
&&\ +2T(1-(\alpha+1)T)\frac{1-e^{-\lambda'(T-\delta)}(1-p)}{1-e^{-\lambda'T}(1-p)}\big]
\ee

Note that $\lambda'sT<\lambda'\tau<<1$, we have
\be
\sum_{s=2}^\alpha(1-sT)p(1-p)e^{-\lambda'sT}&\approx&\sum_{s=2}^\alpha(1-sT)p(1-p)[1-\lambda'sT]\nonumber \\
&\approx&p(1-p)\left[\alpha-1-\frac{\lambda'+1}{2}T(\alpha+2)(\alpha-1)\right] \nonumber \\
&\approx&p(\alpha-1)\label{eq.appendix5_approx_6_1}.
\ee
Moreover, we have the following results on $\frac{1-e^{-\lambda'(T-\delta)}(1-p)}{1-e^{-\lambda'T}(1-p)}$,
\be
\frac{1-e^{-\lambda'(T-\delta)}(1-p)}{1-e^{-\lambda'T}(1-p)}&\approx&\frac{1-\left[1-\lambda'(T-\delta)\right](1-p)}{1-(1-\lambda'T)(1-p))}
\nonumber \\
&\approx&\frac{\lambda'(T-\delta)+p}{\lambda'T+p} \nonumber \\
&=&\frac{T-\delta}{T}+\frac{\delta p}{(\lambda' T+'p)T}\label{eq.appendix5_approx_6_2}
\ee

Substituting (\ref{eq.appendix5_approx_6_1}) and (\ref{eq.appendix5_approx_6_2}) into (\ref{eq.appendix5_var1}), ignoring the second order quantities ($T^2$) that is deemed to be sufficiently small, we have the following result on the variance of $n_s$,
\be
\mathbb D[n_s]\approx \mathbb E[n_s]\left[1+2(\alpha-1)p\right]+2\mathbb E[n_s]^2\left[-(\tau+T/2)+\frac{p\delta}{\lambda' T+p}\right].
\ee
\subsection{Proof of Lemma 1}
For the case of $T>\tau$, we have $N_1=N_0\dff N$, $\frac{P_1}{P_0}=\frac{\hat N_1}{\hat N_0}$, $\frac{\tau'}{\tau'_0}=1$ and the following
\be
D(P_1^B||P_0^B)-D(P_0^B||P_1^B)&=&N\left[(P_0+P_1)\log\frac{P_1}{P_0}+(2-P_1-P_0)\log\frac{1-P_1}{1-P_0}\right]\nonumber \\
&>&N\left[(P_0+P_1)\log\frac{P_1}{P_0}-2\log\frac{1-P_0}{1-P_1}\right] \nonumber \\
&>&N\left[P_1\log\frac{P_1}{P_0}-2\frac{P_1}{1-P_1}\right].
\ee

For $T>\tau$, since $\log\frac{\hat N_1}{\hat N_0}>\frac{2}{1-\hat N_1\tau'}$, it is straightforward that $D(P_1^B||P_0^B)-D(P_0^B||P_1^B)>0$ for the case $T>\tau$.
For $T\le\tau$, we have $\tau'=\tau+\frac{T}{2}$, and the following
\be
D(P_1^B||P_0^B)-D(P_0^B||P_1^B)&\approx&N_0\log\frac{1-P_1}{1-P_0}+N_1P_1\left[\log\frac{P_1}{P_0}-\log\frac{1-P_1}{1-P_0}\right] \nonumber \\
&&\ -N1\log\frac{1-P_0}{1-P_1}-N_0P_0\left[\log\frac{P_0}{P_1}-\log\frac{1-P_0}{1-P_1}\right] \nonumber \\
&=&(N_1P_1+N_0P_0)\frac{P_1}{P_0}-(N_0+N_1-N_1P_1-N_0P_0)\log\frac{1-P_0}{1-P_1} \nonumber \\
&>&N_1\left[P_1\log\frac{P_1}{P_0}-\frac{N_0+N_1}{N_1}\frac{P_1}{1-P_1}\right].
\ee

Noting that $\hat N$ decreases with respect to $\lambda'$, we have
\be
\frac{1}{2\tau'}\le N(\lambda')\le N(0),
\ee
where
\be
N(0)&=&\frac{1}{2\tau'}\frac{1}{1-\left[\frac{\delta}{\tau'}+\frac{(\alpha-1)T}{\tau'}\right]}=\frac{1}{3T}=\frac{1}{2\tau_0'}.
\ee
Thus, we have that $\frac{N_1+N_0}{N_1}<1+\frac{\tau'}{\tau_0'}$. Recall the expressions of $P_0$ and $P_1$ in Theorem~3, we have $\frac{P_1}{P_0}>\frac{\hat N_1}{\hat N_0}$. Moreover, since $\frac{\hat N_1}{\hat N_0}>\frac{1+\frac{\tau'}{\tau'_0}}{1-2\tau'\hat N_1}$, we have that $D(P_1^B||P_0^B)-D(P_0^B||P_1^B)>0$ for the case of $T\le\tau$.
\subsection{Proof of Lemma 3}
We first write the derivative of $N_1\log\frac{1}{1-P_1}$ with respect to $\tau$ in the following,
\be
\frac{\partial N_1\log\frac{1}{1-P_1}}{\partial \tau}&=&\frac{(1-\lambda_1'\tau')\hat N_1}{(1-p_1)\tau'}-\frac{1}{2\tau'^2}\log\frac{1}{1-p_1} \nonumber \\
&=&\frac{p_1(1-\lambda_1'\tau')-(1-p_1)\log\frac{1}{1-p_1}}{2\tau'^2(1-p_1)}.
\ee
Then we need to prove that $p_1(1-\lambda_1'\tau')+(1-p_1)\log{(1-p_1)}\le0$. Noting that we have $(1-x)\log(1-x)\le-x+\frac{x(e^x-1)}{2}$ for $0\le x<0.5$, thus we need to prove the following is satisfied,
\be
p_1(1-\lambda_1'\tau')-p_1+\frac{p_1(e^{p_1}-1)}{2}=\frac{p_1}{2}\left(e^{p_1}-1-2\lambda_1'\tau'\right)\le0,
\ee
which is equivalent to proving the following
\be\label{eq.appendG_ieq1}
\hat N_1\le\frac{\log(1+2\lambda_1'\tau')}{2\tau'}.
\ee
Recall that $\hat N_1=\frac{e^{-\lambda_1'\tau}(1-p)\left[1-e^{-\lambda_1'T}(1-p)\right]}{T}\dff e^{-\lambda_1'\tau}C$, where $C$ is independent of $\tau$. We may need to prove that function $g(\tau)\dff 2(\tau+T/2)e^{-\lambda_1'\tau}C-\log[1+2\lambda_1'(\tau+T/2)]$ decreases strictly with respect to $\tau$, i.e., we need to prove that $\frac{\partial g(\tau)}{\partial \tau}=\frac{2\left[\hat N_1(1-4\lambda_1^2\tau'^2)-\lambda_1'\right]}{1+2\lambda_1'\tau'}\le0$. We first assume it is true, and the strict proof will be given in the rest part of this appendix.

If it is true, the rest work is to prove that the inequality (\ref{eq.appendG_ieq1}) holds when $\tau=T$. Let $z\dff\lambda_1'T$, and $h(z,p)\dff g(T)=3(1-p)e^{-z}\left[1-e^{-z}(1-p)\right]-\log(1+3z)$. Note that for small $z$ and $p$, function $h(z,p)$ increases strictly with respect to $p$. Thus recalling the condition given in Lemma~3, we have
\be
h(z)&=&3(1-p)e^{-z}\left[1-e^{-z}(1-p)\right]-\log(1+3z)\nonumber \\
&\le&3e^{-z-z^3}(1-e^{-z-z^3})-\log(1+3z)\dff r(z).
\ee
For the function $r(z)$, we have
\be
\frac{\partial r(z)}{\partial z}=\frac{3e^{-2(z+z^3)}\left[2(1+3z^2)(1+3z)-e^{z+z^3}(1+3z^2)(1+3z)-e^{2(z+z^3)}\right]}{1+3z}.
\ee
Note that for sufficiently small $z$ we have the following,
\be
&&2(1+3z^2)(1+3z)-e^{z+z^3}(1+3z^2)(1+3z)-e^{2(z+z^3)} \nonumber \\
&&\qquad\qquad\qquad\qquad\qquad \le 2(1+3z^2)(1+3z)-(1+z+z^3)(1+3z^2)(1+3z)-(1+z+z^3)^2\nonumber \\
&&\qquad\qquad\qquad\qquad\qquad =-z^2(1-3z+14z^2+3z^3+10z^4)\le0,
\ee
where the last inequality holds due to the small $z$.
Thus we have $\frac{\partial r(z)}{\partial z}<0$, which leads to $g(\tau)\le g(T)=h(z)\le r(z)\le r(0)=0$.

Finally we prove that $\frac{\partial g(\tau)}{\partial \tau}<0$ in the following. Recall that the inequality $h(z)\le0$ have been proved, we have
\be
\hat N_1&\le&\frac{3(1-p)e^{-z}\left[1-e^{-z}(1-p)\right]}{T} \nonumber \\
&\le&\frac{\log(1+3z)}{3T}\le z/T=\lambda_1',
\ee
and thus $\hat N_1(1-4\lambda_1^2\tau'^2)-\lambda_1\le\hat N_1-\lambda_1'\le0$, which shows that $\frac{\partial q(\tau)}{\partial \tau} \leq 0$, i.e., $g(\tau)$ is decreasing with respect to $\tau$.

\subsection{Proof of Lemma 4}
We write the derivative of $r(\tau)$ with respect to $\tau$ as follows,
\be
\frac{\partial r(\tau)}{\partial \tau}&=&\frac{1}{P_0}\frac{\partial P_0}{\partial \tau}-\frac{1}{(1-P_1)P_1}\frac{\partial P_1}{\partial \tau} \nonumber \\
&=&\frac{1-\lambda_0'\tau'-\frac{p(1-\lambda_0'\delta)}{p+\lambda_0'T}}{\tau'\left[1-\left(\frac{p\delta}{\tau'(\lambda_0' T+p)}+\frac{(\alpha-1)p}{\hat N\tau'}\right)\right]}-\frac{1-\lambda_1'\tau}{\tau'(1-2\tau'\hat N_1)}.
\ee
We first prove that $\frac{1}{P_0}\frac{\partial P_0}{\partial \tau}$ is less than $\frac{1}{\tau'}$. Note that
\be
\hat N_0 T&=&e^{-\lambda_0'\tau}(1-p)\left[1-e^{-\lambda_0'T}(1-p)\right] \nonumber \\
&=&e^{-\lambda_0'(\tau+T)}(1-p)\left[e^{\lambda_0'T}-(1-p)\right] \nonumber \\
&\ge&e^{-\lambda_0'(\tau+T)}(1-p)(\lambda_0'T+p).
\ee
Substituting the condition into the above inequality, we have
\be\label{eq.lemma4_ieq1}
\frac{\hat N_0}{\lambda_0'T+p}\ge\frac{\alpha-1}{(\alpha+\frac{1}{2})T}=\frac{\alpha-1}{\tau'-\delta}.
\ee
Note that
\be
&&1-\left(\frac{p\delta}{\tau'(\lambda_0' T+p)}+\frac{(\alpha-1)p}{\hat N\tau'}\right)-\left[1-\lambda_0'\tau'-\frac{p(1-\lambda_0'\delta)}{p+\lambda_0'T}\right] \nonumber \\
&&\quad\quad\ge\frac{p}{\tau'}\left[\frac{\tau'-\delta}{p+\lambda_0'T}-\frac{\alpha-1}{\hat N_0}\right]\ge 0,
\ee
where the final inequality is obtained based on (\ref{eq.lemma4_ieq1}). Then it can be seen that $\frac{1}{P_0}\frac{\partial P_0}{\partial \tau}\le\frac{1}{\tau'}$.

According to the assumption of sufficiently small $\lambda_1'\tau'$, we have that $\hat N_1\ge\frac{e^{-\lambda_1'\tau}(1-e^{-\lambda_1'T})}{T}\ge\frac{e^{-\lambda_1(\tau+T)}\lambda_1'T}{T}>\frac{1}{2}\lambda_1'$. Then it can be seen that $\frac{1}{(1-P_1)P_1}\frac{\partial P_1}{\partial \tau}=\frac{1-\lambda_1'\tau}{\tau'(1-2\tau'\hat N_1)}>\frac{1}{\tau'}$.

Based on the above analyses, it can be proved that $\frac{\partial r(\tau)}{\partial \tau}$ is negative.

\bibliographystyle{./IEEEtran}
\bibliography{./mybib}
\end{document}